\documentclass[10pt,twocolumn,aps,prl,nofootinbib]{revtex4-1}

\usepackage{amsmath}
\usepackage{graphicx}
\usepackage[utf8]{inputenc}

\usepackage[colorlinks,pdfusetitle]{hyperref}
\hypersetup{allcolors=[rgb]{1,0.56,0}}

\newcommand{\orcid}[1]{\href{https://orcid.org/#1}{#1}}
\newcommand{\e}[1]{\times10^{#1}}

\begin{document}

\title{Ultra-High-Energy Tau Neutrino Cross Sections with GRAND and POEMMA}

\author{Peter B.~Denton}
\email{pdenton@bnl.gov}
\thanks{\orcid{0000-0002-5209-872X}}

\author{Yves Kini}
\email{kiniyves@gmail.com}
\thanks{\orcid{0000-0002-0428-8430}}

\affiliation{High Energy Theory Group, Physics Department, Brookhaven National Laboratory, Upton, NY 11973, USA}

\hyphenation{UHECRs}

\date{\today}

\begin{abstract}
Next generation neutrino experiments will push the limits in our understanding of astroparticle physics in the neutrino sector to energies orders of magnitude higher than the current state-of-the-art high-energy neutrino experiment, IceCube.
These experiments will use neutrinos to tell us about the most extreme environments in the universe, while simultaneously leveraging these extreme environments as probes of neutrino properties at the highest energies accessible in the foreseeable future: $E\sim10^9$ GeV.
At these energies neutrinos are readily absorbed in the Earth.
Assuming an isotropic distribution, by looking at how the flux varies as a function of angle through the Earth, we show that it is possible to extract the $\nu_\tau$-$N$ cross section with precision at the $\sim20\%$ level ($1\sigma$ assuming Wilks' theorem) given $N_{\rm events}\sim100$ events.
\end{abstract}

\maketitle

\section{Introduction}
The origins of ultra-high-energy cosmic rays (UHECRs) have been one of biggest mysteries in modern astrophysics.
Discovering their sources will provide crucial information on where they are produced in the universe and how they are accelerated to such high energy.
One way to probe this enigma is to detect neutrinos coming from the interaction of UHECRs and photons from the cosmic microwave background.
Unlike UHECRs, neutrinos are not deflected in magnetic fields and the universe is much more transparent to neutrinos \cite{Weiler:1982qy} making them an excellent orthogonal probe to understand the nature of the extreme sources accelerating UHECRs.

\begin{figure}
\centering
\includegraphics[width=\columnwidth]{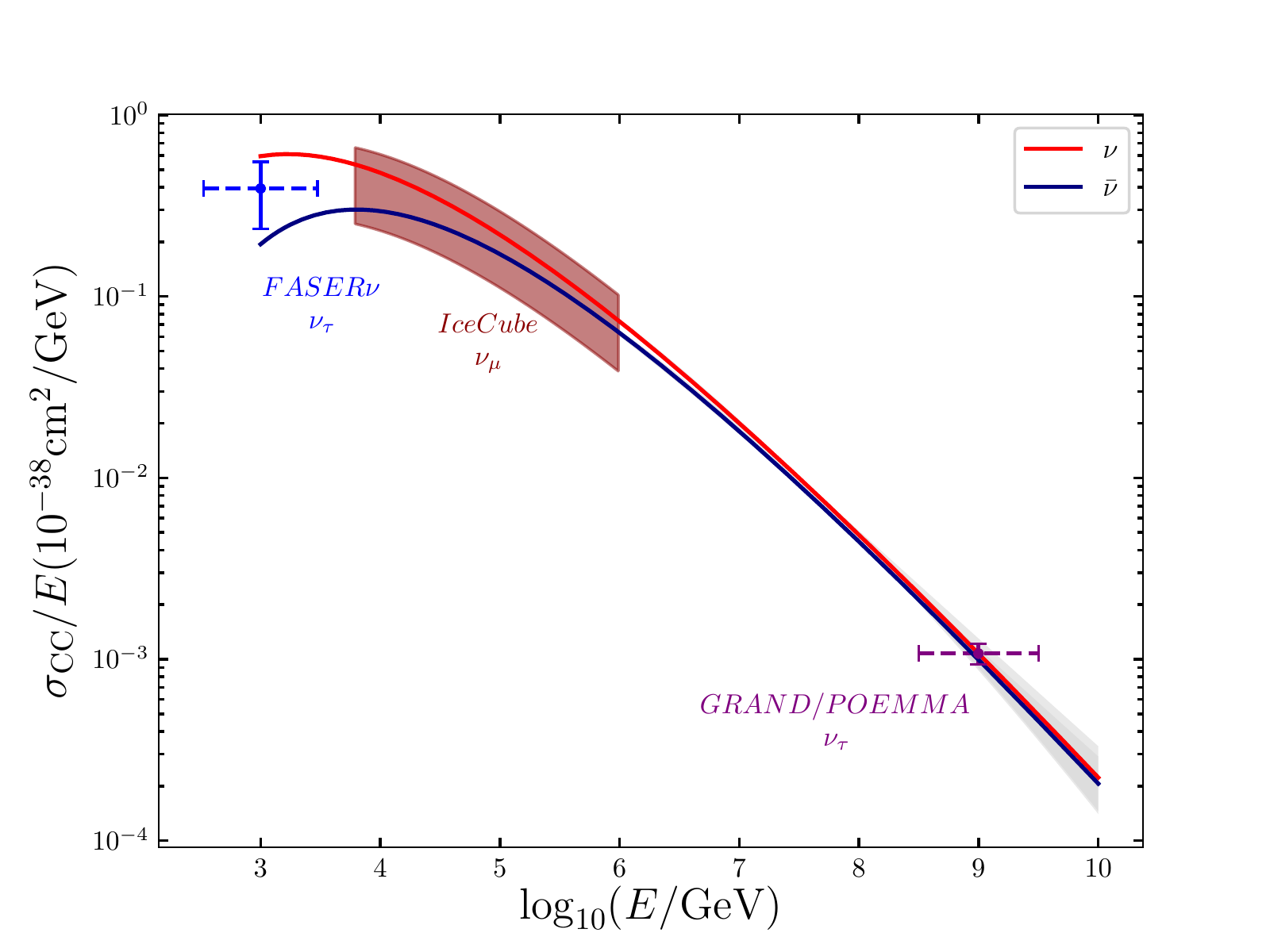}
\caption{Neutrino-nucleon (red) and antineutrino-nucleon (blue) cross section for charged-current (CC) interactions including PDF uncertainties in gray from \cite{Connolly:2011vc}.
The existing constraint on the $\nu_\mu$ cross section from IceCube is shown as a red band \cite{Aartsen:2017kpd}.
FASERnu at the LHC is expected to measure the $\nu/\bar\nu$ weighted energy-dependent cross section at $E\sim1$ TeV \cite{Abreu:2019yak}.
Other laboratory measurements exist at lower energies.
The expected sensitivity by GRAND or POEMMA, as determined in this paper, is shown in purple assuming 100 events measured.
Horizontal error bars approximate the energy range over which the measurements are expected to cover, not the energy resolution.}
\label{fig:cross_section}
\end{figure}

On the other hand, this guaranteed source of neutrinos provides an excellent opportunity to test the Standard Model (SM) of particle physics and probe the nature of neutrinos at ultra-high energies (UHE) $E\gtrsim10^9$ GeV; for a recent review of new physics tests at upcoming neutrino experiments see ref.~\cite{Arguelles:2019xgp}.
One key test of neutrino properties at high energies is to determine if the neutrino-nucleon cross section behaves as expected.
To date neutrino-nucleon cross sections have only been measured in laboratory environments up to $E\sim350$ GeV \cite{Tanabashi:2018oca,Tzanov:2005kr}.
Upcoming experiments like FASER$\nu$ at the LHC will measure neutrino-nucleon cross sections of each flavor at $E\sim10^3$ GeV with the excellent precision for $\nu_e$ and $\nu_\mu$ but only $\sim40\%$ precision for $\nu_\tau$ \cite{Feng:2017uoz,Abreu:2019yak,Abreu:2020ddv}.
By measuring the absorption rate in the Earth, IceCube has determined that the neutrino-nucleon cross section is compatible with the SM at the $\sim50\%$ level at $1\sigma$ in the $10^4$ GeV $\lesssim E\lesssim10^6$ GeV range \cite{Aartsen:2017kpd,Bustamante:2017xuy,Anchordoqui:2019ufu}.
These $E\gtrsim1$ TeV sensitivities and measurements including the result from this paper are shown in fig.~\ref{fig:cross_section}.

All measurements to date are consistent with the theoretical predictions.
The theory predictions are quite precise up to $E\sim10^8$ GeV at which point the predictions lose precision due to limitations in extrapolating parton distribution functions (PDFs) at low Bjorken-$x$ \cite{CooperSarkar:2011pa,Connolly:2011vc,Arguelles:2015wba,Garcia:2020jwr,Bertone:2018dse,Chen:2013dza}.
In addition to UHE neutrino experiments, measurements from the LHC and other current and future laboratory accelerator experiments can improve these PDFs as well.
In addition to possibly constraining PDFs at smaller $x$ than can be probed by upcoming collider experiments \cite{Hobbs:2019sut}, UHE neutrinos can also provide important constraints on nuclear PDFs \cite{Garcia:2020jwr}.
On the new physics side, there are various scenarios that predict significant increases or decreases to the total cross section such as large extra dimensions \cite{AlvarezMuniz:2001mk}, sphalerons \cite{Ellis:2016dgb}, or color glass condensate \cite{Henley:2005ms}.

Several current and next-generation neutrino experiments have sensitivity to UHE neutrinos at $E\sim10^9$ GeV.
While there is a guaranteed flux of cosmogenic neutrinos thanks to UHECRs scattering off the cosmic microwave background \cite{Greisen:1966jv,Zatsepin:1966jv}, there may also be an additional component from galaxy clusters, pulsars, active galactic nuclei, and gamma ray bursts among other possible accelerators \cite{Stecker:1991vm,Waxman:1998yy,Murase:2006mm,Murase:2007yt,Murase:2008mr,Murase:2008yt,Kotera:2009ms,Murase:2009pg,He:2012tq,Fang:2013vla,Murase:2014foa,Bustamante:2014oka,Murase:2015ndr,Bustamante:2016wpu,Denton:2017jwk,Fang:2017zjf,Rodrigues:2020pli,Righi:2020ufi} that may or may not be an extension of the flux IceCube has measured \cite{Aartsen:2013jdh}.
The Antarctic Impulsive Transient Antenna (ANITA) \cite{Barwick:2005hn}, IceCube \cite{Aartsen:2016ngq}, and the Pierre Auger Observatory \cite{Aab:2015kma} have already placed constraints on ultra-high-energy neutrinos.
The proposed/under construction Giant Radio Array for Neutrino Detection (GRAND) \cite{Alvarez-Muniz:2018bhp} and the Probe Of Extreme Multi-Messenger Astrophysics (POEMMA) \cite{Olinto:2017xbi} have good sensitivity to most of the parameter space of the expected flux of cosmogenic neutrinos \cite{AlvesBatista:2018zui,Moller:2018isk}.
In addition to the experiments discussed above there are several other proposed techniques to detect UHE neutrinos including techniques involving active radar or optical detectors \cite{Aartsen:2014njl,Romero-Wolf:2014pua,Nam:2015vak,Vieregg:2015baa,Sasaki:2017uta,Otte:2018uxj,Prohira:2019glh,Wissel:2019alx,Wissel:2020sec,Romero-Wolf:2020pzh}.

UHE neutrino experiments are dominantly sensitive to tau neutrinos ($\nu_\tau$).
This unique sensitivity exists because a UHE $\nu_\tau$ will travel through the Earth and then interact with a mean free path near the surface of the Earth of $\lambda\sim1000$ km at $E\sim10^9$ GeV.
If the interaction is neutral-current (NC) it will lose some energy and continue propagating.
If it is charged-current (CC) then a tau lepton ($\tau$) will be produced.
The $\tau$ will then lose energy in matter before decaying.
If it decays in matter the process will continue, albeit at lower energies, since one of the decay products is always a $\nu_\tau$; this mechanism is known as $\nu_\tau$ regeneration \cite{Halzen:1998be,Dutta:2002zc,Bigas:2008sw}.
If the $\tau$ escapes the Earth it will decay in the atmosphere.
Most of these decays will result in a large air shower\footnote{A $\tau$ decays to a muon and two neutrinos 17\% of the time \cite{Tanabashi:2018oca} which will not result in an easily detectable air shower.} which can then be detected with various different detection technologies.

Alternatively, if one observes an air shower coming up out of the Earth, it \emph{must} be due to a $\nu_\tau$ propagating in the Earth which experiences a CC interaction (at least one) producing a $\tau$ which then escapes the Earth and then decays.
There is no other process in the SM that will lead to such a signature.
Thus the Earth provides a sort of filter to block all cosmic rays and only permits neutrinos through.

Given the significant absorption rate of UHE neutrinos, this leads to a suppression of the flux depending on the amount of Earth through which the neutrinos traverse.
This means that by measuring the angular distribution and comparing with the local topology and the curvature of the Earth, one can extract the absorption rate and, given an estimate of the weak charge density of the Earth in different layers, the neutrino-nucleon cross section.
In addition, if the spectrum continues to even higher energies, horizontal trajectories where neutrinos interact in the atmosphere may become dominant \cite{Kusenko:2001gj,PalomaresRuiz:2005xw}.
We focus on neutrinos interacting in solid matter only.

In this paper we will use the sensitivity of GRAND and POEMMA to determine the tau neutrino-nucleon cross section.
First, we discuss our simulation of $\nu_\tau$ propagation and the relevant experimental details.
We then present our numerical results.
Finally, we discuss some interesting aspects of the results and plans going forward and conclude.

\section{Tau Neutrino Simulation}
\label{sec:nu tau simulation}
\begin{figure}
\centering
\includegraphics[width=\columnwidth]{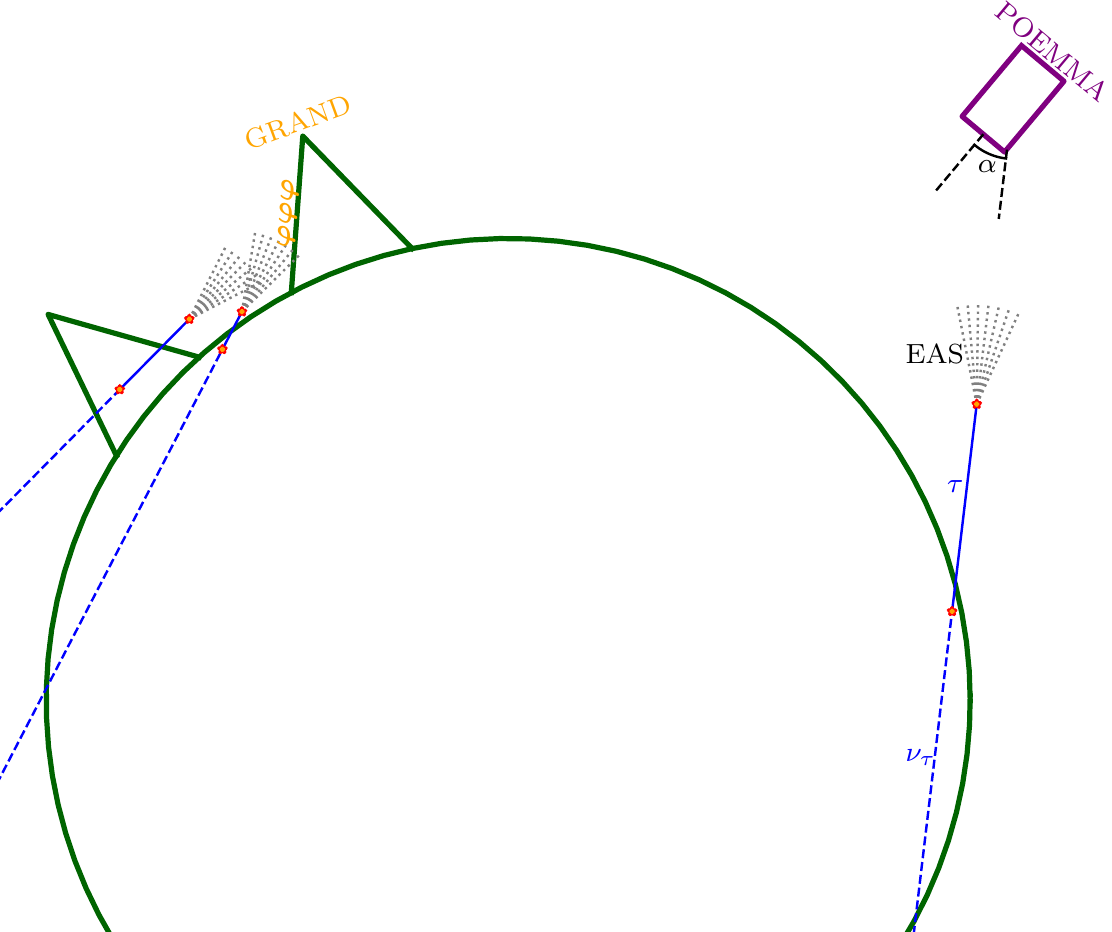}
\caption{A schematic (not to scale) representation of $\nu_\tau$ trajectories.
First, a $\nu_\tau$ passes through a mountain or the Earth and interacts near the surface producing a $\tau$.
The $\tau$ then escapes the Earth and decays hadronically into an extensive air shower (EAS) which can be measured by GRAND or POEMMA.
The angle $\alpha$ denotes the angle between the shower trajectory and the center of the Earth.}
\label{fig:schematic}
\end{figure}

To handle the $\nu_\tau$ propagation in matter including regeneration effects, we use the publicly available \texttt{NuTauSim} software \cite{Alvarez-Muniz:2017mpk,nutausim}, see also \cite{Safa:2019ege}.
For our fiducial cross section and tau energy loss models we use the central values from Connoly, et al.~\cite{Connolly:2011vc} and Abramowicz, et al.~\cite{Abramowicz:1997ms} respectively.
We made several modifications to the code based on the unique topology of the surface of the Earth relevant for GRAND as shown schematically in fig.~\ref{fig:schematic}.

We consider two experimental configurations, one for GRAND and one for POEMMA; other UHE neutrino experiments should have similar sensitivities to the cross section depending primarily on the statistics achieved.
The UHE neutrino flux is extremely uncertain; the guaranteed component from UHECRs varies by about two orders of magnitude depending on the redshift evolution of the UHECR sources and the UHECR mass composition \cite{AlvesBatista:2018zui,Moller:2018isk}.
In addition there may be an additional component of UHE neutrinos produced at sources that is largely unconstrained \cite{Stecker:1991vm,Waxman:1998yy,Murase:2006mm,Murase:2007yt,Murase:2008mr,Murase:2008yt,Kotera:2009ms,Murase:2009pg,He:2012tq,Fang:2013vla,Murase:2014foa,Bustamante:2014oka,Murase:2015ndr,Bustamante:2016wpu,Denton:2017jwk,Fang:2017zjf,Rodrigues:2020pli,Righi:2020ufi}.
As a benchmark we assume that an experiment will observe $N_{\rm events}=100$ events\footnote{We discuss neutrinos detected instead of neutrino flux since the exact exposures and efficiencies of different experiments are still being determined and the number of events is the primary parameter for determining the precision with which the cross section can be determined.} assuming the SM cross section.
This number is plausible as the expected event rate in GRAND from the cosmogenic component \emph{alone} is 1-18 events per year depending on how optimistic or pessimistic the UHECR parameters are for the resultant UHE neutrino flux and POEMMA has a comparable sensitivity \cite{Alvarez-Muniz:2018bhp}; additional components of the flux that may well exist will increase this.

To parameterize deviations from the expected neutrino-nucleon cross section, we introduce an energy-independent scale parameter $S$ that rescales the entire $\nu-N$ cross section (CC and NC together) which is the same approach used in ref.~\cite{Aartsen:2017kpd}, 
\begin{equation}
\sigma=S\sigma_{\rm SM}\,.
\end{equation}
That is, $S=1$ is the SM expectation.
When comparing different values of $S$, we assume that the initial neutrino flux and detector exposure are the same, so $N_{\rm events}$ changes.
In principle one could also examine the energy dependent cross section as well depending on the statistics and the energy resolution of the detector.
While changing the cross section is not exactly the same as changing the mean free path due to $\nu_\tau$ regeneration, the effect of regeneration is small as most of the air showers detected will be from events that experienced a single interaction.

Both GRAND and POEMMA will have good sensitivity to neutrinos above few$\e7$ GeV by measuring the radio signal (GRAND) or the Fluorescence and Cherenkov light (POEMMA) from air showers \cite{Martineau-Huynh:2015hae,Anchordoqui:2019omw}.
Therefore, we have set the minimum $\tau$ energy to $4\e7$ GeV during propagation in the Earth ensuring that the resultant shower has energy $\gtrsim$ few$\times10^7$ GeV.

GRAND will cover an area of 200,000 km$^2$ with radio antennae to detect horizontal air showers coming from either mountain-passing or Earth-skimming $\nu_\tau$ events.
Since the exact location for GRAND is still being determined, we approximate the local topography as a detector that is on average 2 km in elevation on the side of a mountain facing another mountain 10 km away that is 6 km tall and 100 km wide with a density $\rho=2.9$ g/cc \cite{Alvarez-Muniz:2018bhp}.
The horizon is then at $\alpha=88.6^\circ$ where the angle $\alpha$ is defined as 180$^\circ-\theta_z$ where $\theta_z$ is the zenith angle for the detector at a height 2 km and for a given neutrino trajectory to the detector.
The specific topography would have to be accounted for once GRAND starts detecting neutrinos, but this simplified model should demonstrate the impact of mountains on the cross section sensitivity.
Therefore neutrinos can arrive at GRAND after passing through 1) the (spherical) Earth, 2) the mountain, 3) both the Earth and the mountain, as well as 4) the Earth, the air, and then the mountain.
While the last two options represent only a small solid angle, we account for each of these different paths.
We bin the data in 0.5$^\circ$ width bins based on estimates from GRAND.
Although GRAND could potentially reach 0.1$^\circ$ resolution, we have checked that the impact on the cross section sensitivity is not too large.

POEMMA will orbit the Earth at varying altitudes ranging from 525 km to 1,000 km.
We assume a fixed altitude of 525 km and angular resolution of 1$^\circ$ \cite{Olinto:2017xbi}.
Thus the horizon is at $\alpha=67.5^\circ$.
In both cases we model the Earth density with the preliminary reference earth model \cite{Dziewonski:1981xy}.
While newer Earth models have more detailed crust descriptions, the fluctuations within the crust among these for most of the crust is at the $\sim2\%$ level (see e.g.~ref.~\cite{Bakhti:2020tcj} for a comparison of several models) which is negligible given the anticipated statistics.
Our main analyses do not include a water layer; it has been noted that such a layer increases the event rate relative to rock only \cite{PalomaresRuiz:2005xw}.
We have verified that our cross section results for POEMMA, for which this could potentially make a difference, are unaltered by the inclusion of such a layer.

\section{Results}
\label{sec:results}

\begin{figure*}
\centering
\includegraphics[width=\columnwidth]{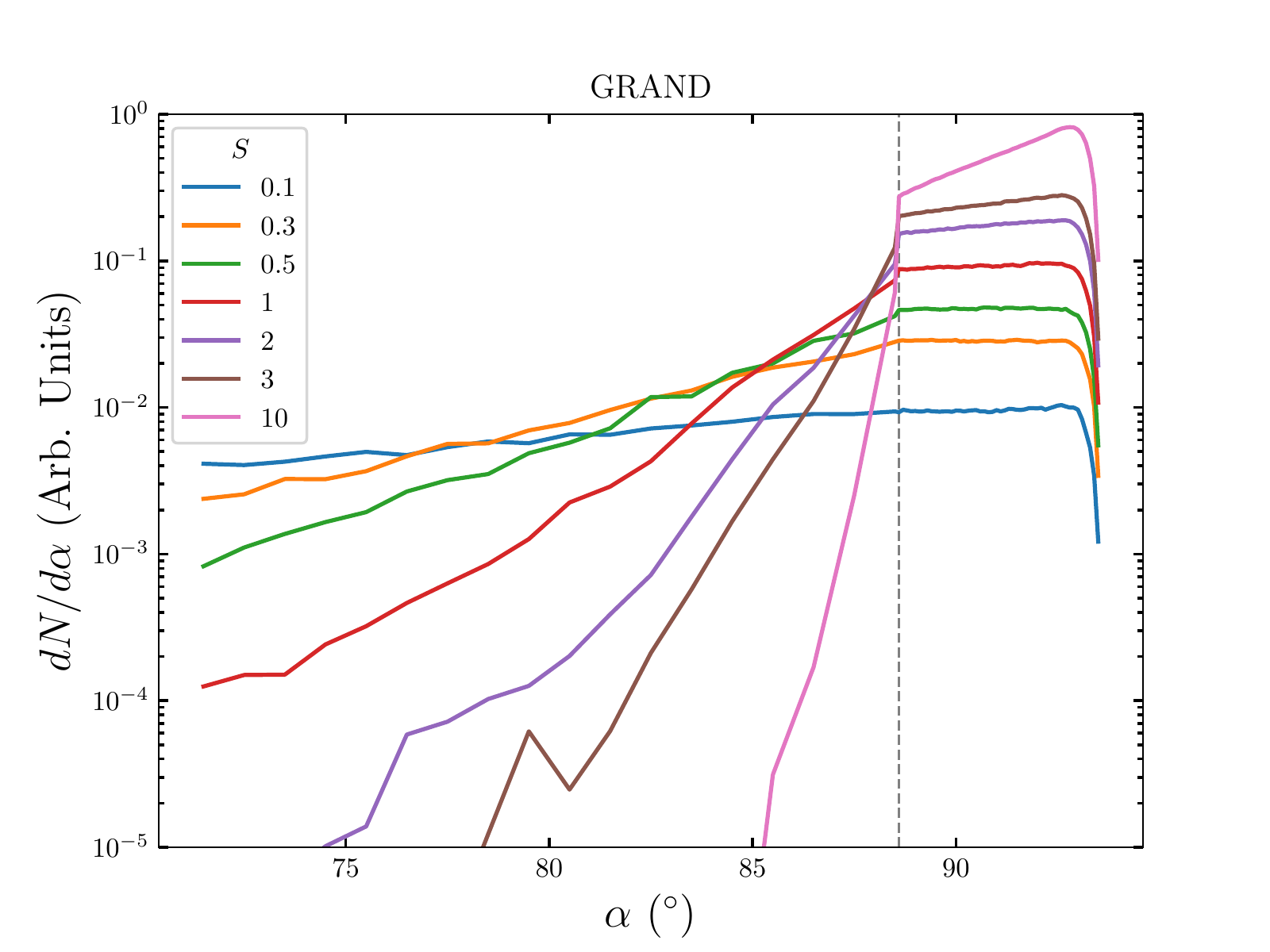}
\includegraphics[width=\columnwidth]{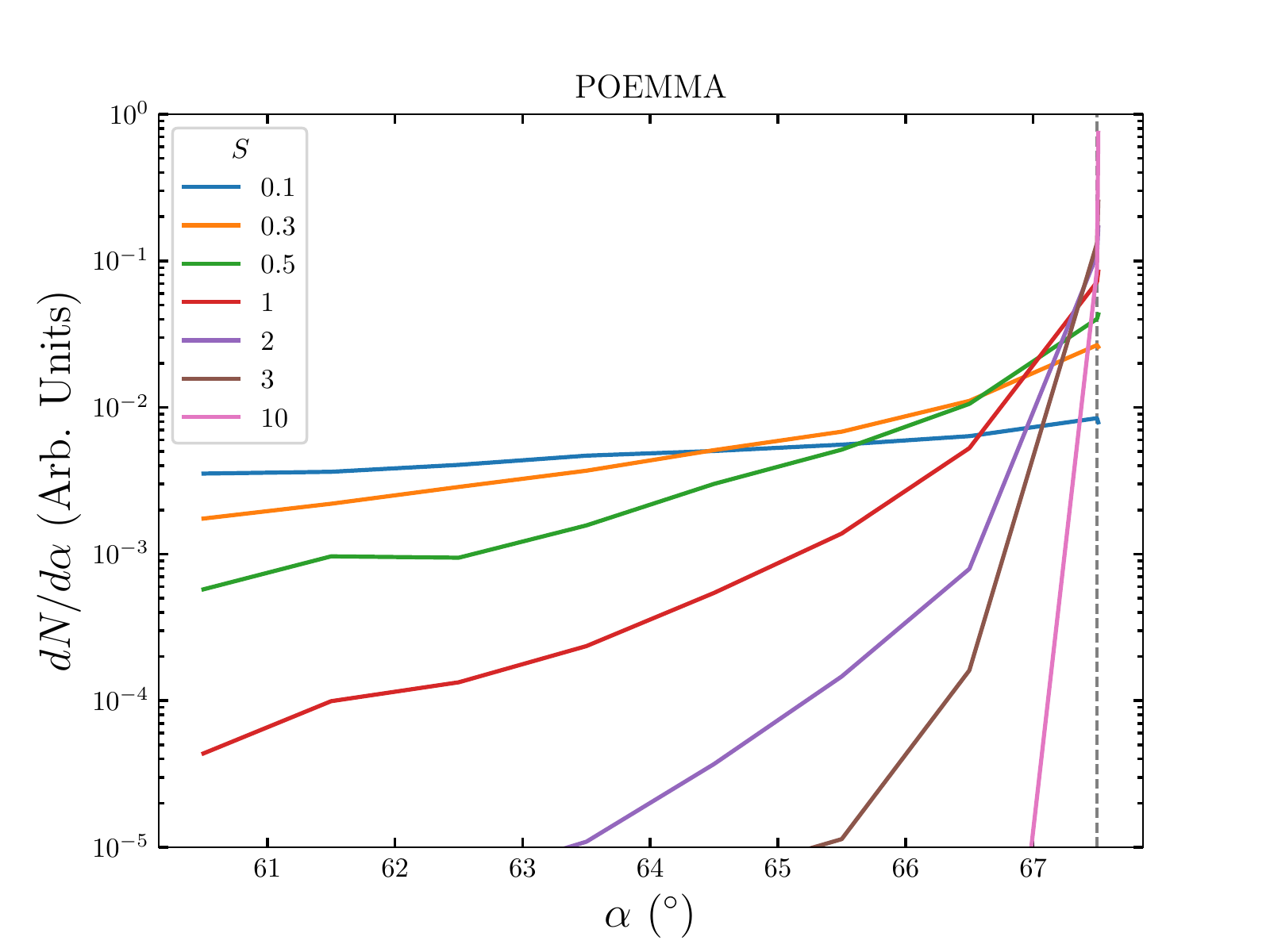}
\caption{The detected angular distribution of the flux for GRAND (left) and POEMMA (right) at $E=10^9$ GeV.
The different curves refer to different cross section scalings with $S=1$ being the SM.
The horizons, at $88.6^\circ$ and and $67.5^\circ$ for GRAND and POEMMA respectively, are shown with vertical dashed lines.
Detections at GRAND with $\alpha>88.6^\circ$ come from the interactions which took place in the mountain.}
\label{fig:number_events}
\end{figure*}

In fig.~\ref{fig:number_events}, we show the expected angular distributions for both GRAND and POEMMA for various cross section scale factors $S$.
In the case of GRAND, the opposite mountain considerably increases the number of events.
In addition, given the size of the mountain, it is clear how the slope of the event rate varies depending on the cross section providing a powerful tool for determining the cross section; for large cross sections the slope is quite steep, while for smaller cross sections the slope is nearly flat.

Next, as a test statistic, we calculate the $\chi^2$ function between a given cross section and $S=1$.
In a given angular bin we have,
\begin{multline}
\chi^2(S,\alpha_i,\beta)=2\left[\vphantom{\left(\frac{\beta N_i}{N_i}\right)}(1+\beta)N_i(S)-N_i(1)\right.\\
\left.+N_i(1)\log\left(\frac{N_i(1)}{(1+\beta)N_i(S)}\right)\right]\,,
\label{eq:chisq1}
\end{multline}
where $N_i(S)$ is the number of events detected in $\alpha$ bin $i$, $\beta$ is the normalization pull term, and for cross section scaled by $S$.
Then the total $\chi^2$ is
\begin{equation}
\chi^2(S)=\min_\beta\sum_i\chi^2(S,\alpha_i,\beta)\,.
\label{eq:chisq2}
\end{equation}
We take the sum over angles down to 20$^\circ$ and 5$^\circ$ below the horizon for GRAND and POEMMA respectively, beyond which points the statistics considerably fall off; we have verified that extending these ranges further does not affect our results.
We include a marginalization over the normalization $\beta$
left to freely float
since we do not know the true flux and changing the cross section, to leading order, appears simply as a change in the total number of events.
This ensures that we are only probing the effect due to the changing angular distribution which appears at higher order\footnote{In principle one could apply a prior based on the estimated uncertainty of the cosmogenic flux which would slightly improve our results, but would not be robust.}.
The $\chi^2$ curves are shown in the appendix along with a discussion of the impact of fixing the initial neutrino energy to $E=10^9$ GeV.
We find that at $\Delta\chi^2=1$ GRAND or POEMMA with 100 events can constrain the neutrino-nucleon cross section to about 20\% precision at $E\sim10^9$ GeV.
This maps on to the 1 $\sigma$ level if Wilks' theorem is satisfied, although given the low statistics per angular bin, a more careful statistical analysis would be required given real data.

\begin{figure}
\centering
\includegraphics[width=\columnwidth]{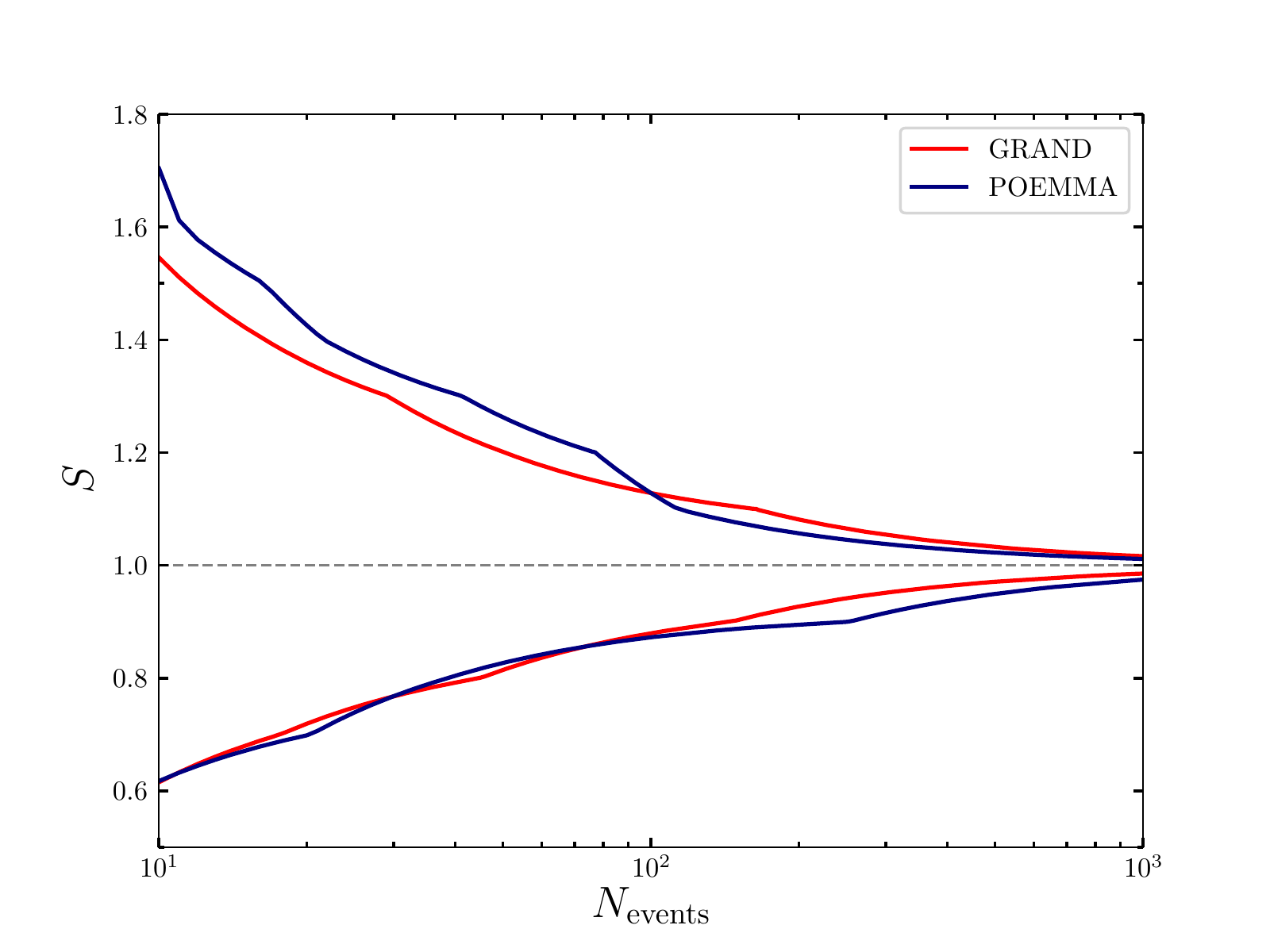}
\caption{Dependence of the cross section uncertainty at $\Delta\chi^2=1$ on the number of events for $E=10^9$ GeV.}
\label{fig:scale_vs_events}
\end{figure}

Finally, the impact of statistics on the cross section sensitivity as shown in fig.~\ref{fig:scale_vs_events}.
We see that for a fixed amount of statistics, each of GRAND and POEMMA has a comparable level of sensitivity.

\section{Discussion}
\label{sec:discussion}
Throughout this analysis we assumed that the incoming neutrino flux is isotropic.
If the flux carries some structure that correlates with the exposure of the experiments then a possible degeneracy between the cross section and the anisotropy could exist, although given the unique exposure of each experiment, such a degeneracy is unlikely.
It is known that the UHECR flux is quite isotropic \cite{Ahlers:2017wpb,Aab:2018chp,Abbasi:2018tqo,Aab:2020xgf,Abbasi:2020fxl,Aab:2020mfn} and the cosmogenic neutrino flux is expected to be even more isotropic as it is likely coming from a broader redshift distribution which would further weaken any anisotropies present in the UHECR flux due to local structure.
In addition, the neutrino flux measured by IceCube at $100$ TeV $\lesssim E\lesssim$ 1 PeV does not correlate with the galaxy \cite{Ahlers:2015moa,Denton:2017csz,Aartsen:2017ujz} and if the flux they have measured continues up to these energies, a Galactic contribution becomes less likely as the energy increases.
If a UHE neutrino point source is identified with multiple events, this cross section measurement can still be performed as the point source will appear at a different angle between the detector and the Earth at different times.

For experiments like IceCube and FASER$\nu$ (and even more so for those experiments at lower energies) neutrinos and anti-neutrinos need to be considered separately.
At the energies that GRAND and POEMMA are sensitive to $\sigma_{\nu N}=\sigma_{\bar\nu N}$ to a good approximation.
This is due to the fact that at high energies protons and anti-protons and indistinguishable as the valence quark contributions become negligible due to Pomeranchuk's theorem.

\section{Conclusions}
\label{sec:conclusions}
After the many successes of IceCube including the measurement of the extragalactic high-energy neutrino flux up to $\mathcal O($few$)$ PeV, there is now a serious effort
around the globe
to develop technology to probe neutrino physics at the EeV scale.
These upcoming experiments will have a rich physics program including much of the same astroparticle physics as IceCube is already sensitive to plus the addition of the cosmogenic neutrino flux and connections to ultra-high-energy cosmic rays.
Beyond that, these upcoming experiments will be able to probe neutrino particle physics at the highest energies probably ever accessible.

In this paper we have highlighted one such example: the tau neutrino-nucleon cross section at $E\sim1$ EeV.
At these energies the cross section is becoming uncertain due to PDF uncertainties and can also provide a probe of various new physics scenarios.
While the flux is very uncertain, we have estimated the expected level of precision with which GRAND and POEMMA can be expected to constrain the cross section for various numbers of events.
In the scenario where 100 events are detected, we find that both GRAND and POEMMA can get $\sim20\%$ precision and the impact of statistics is shown in fig.~\ref{fig:scale_vs_events}.
In addition, in the event that multiple such ultra-high-energy neutrino experiments are constructed, they can perform combined analyses to further enhance their statistical reach.
We hope that this study opens up the possibility to performing additional particle physics tests of ultra-high-energy neutrinos.
Finally, while the tau neutrino is generally the poorest measured particle in the Standard Model, this measurement would change that such that, at least at ultra-high energies, it would be better measured than the other two neutrino flavors due to its unique detection signature.

\begin{acknowledgments}
We thank Mauricio Bustamante and Sergio Palomares-Ruiz for helpful comments.
PBD acknowledges the United States Department of Energy under Grant Contract desc0012704.
YK wishes to acknowledge the African School for Fundamental Physics and Applications (ASP) supported by multiple international institutes and organizations in Africa, Asia, Europe, and the USA.
\end{acknowledgments}

\appendix

\section{Appendix: Energy Dependence}
Throughout the main analysis we assumed that the neutrino flux only contains neutrinos with initial energies of $E=10^9$ GeV.
We now justify this assumption.

\begin{figure*}
\centering
\includegraphics[width=0.49\textwidth]{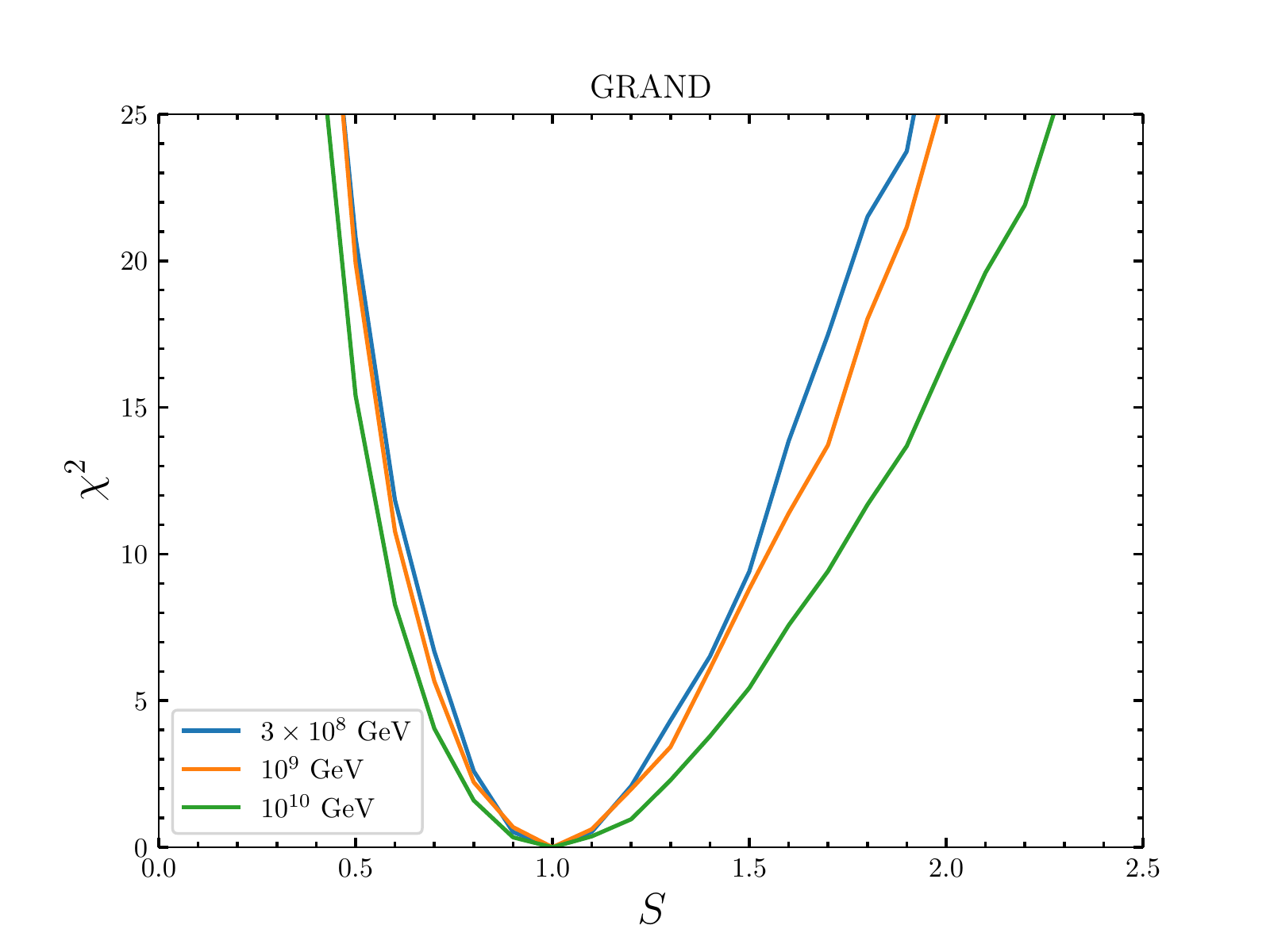}
\includegraphics[width=0.49\textwidth]{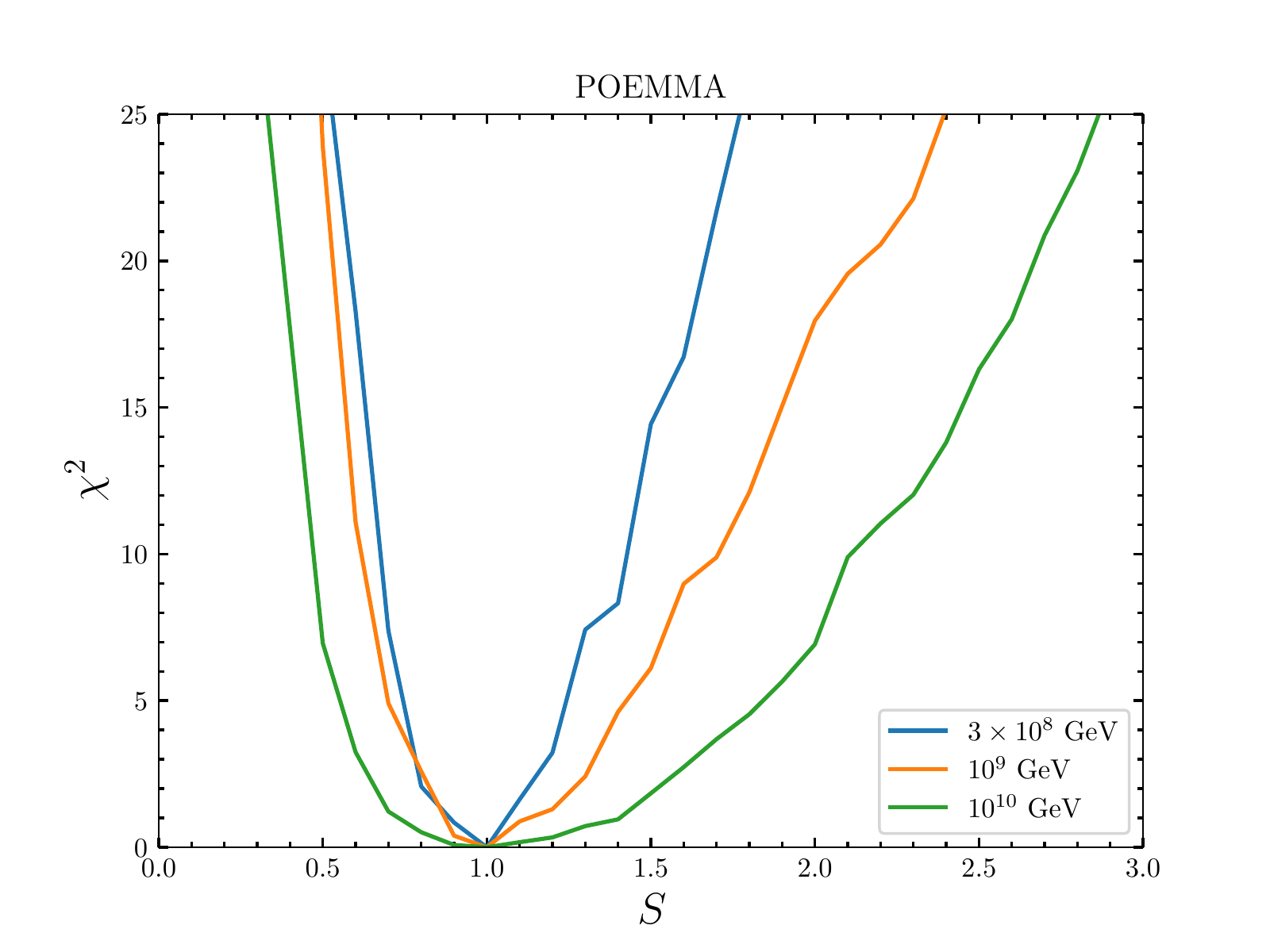}
\caption{The $\chi^2$ function as a function of the cross section scaling for different energies assuming that the flux is such that the experiment measures 100 events for $S=1$. The curves were obtained assuming neutrino interactions from the tip of the mountain up to $\approx 20^\circ$ below the horizon for GRAND while for POEMMA we went $\approx 5^\circ$ below the horizon. We see that the energy impact depends on the experiment and is slightly asymmetric in $S$.}
\label{fig:chisq}
\end{figure*}

First, the cosmogenic flux is expected to peak around $E\approx10^8$-$10^9$ GeV \cite{AlvesBatista:2018zui} although there could be an additional component to the UHE neutrino flux that goes to higher energies.
Second, GRAND's sensitivity peaks around $E\approx10^{8.5}$-$10^9$ GeV \cite{Alvarez-Muniz:2018bhp} and the POEMMA sensitivity is similar or a bit higher \cite{Olinto:2017xbi}.
Third, the energy resolution of the air showers is only modest, at the $\sim25\%$ level \cite{Anchordoqui:2019omw}.
Fourth, the shower energy is not the same as the initial neutrino energy: even if a $\nu_\tau$ only experiences one interaction in the Earth, the $\tau$ will lose energy in the Earth before escaping.
In addition, when it decays, its decay products always include a $\nu_\tau$ which carries away some of the energy invisibly.

Thus estimating the true neutrino energy requires unfolding the true neutrino energy spectrum from the observed air shower spectrum accounting for the details mentioned above including a parameterization of the true neutrino flux; there will likely be differences depending on whether only the cosmogenic flux is assumed or if a power low component is also included.
These difficulties, combined with the low to moderate statistics expected, imply that a single energy bin is a reasonable assumption for the cross section sensitivity.
In fig.~\ref{fig:cross_section} we estimated the energy range over which the cross section might be measured by considering the width of the sensitivity of GRAND and POEMMA.
Once data is acquired, it may turn out that the true spectrum is broader or narrower, or centered at higher or lower energies, but the error bar shown represents a reasonable estimate based on the sensitivities of the experiments.

We then checked the impact of changing the energy of the neutrinos from $10^9$ GeV to various other energies in fig.~\ref{fig:chisq}.
We found that the impact on the precision to be quite modest, thus focusing on $10^9$ GeV alone as opposed to a more realistic spectrum and detector efficiency should have a small impact on the true sensitivity.

\bibliography{UHEnus}

%merlin.mbs apsrev4-1.bst 2010-07-25 4.21a (PWD, AO, DPC) hacked
%Control: key (0)
%Control: author (8) initials jnrlst
%Control: editor formatted (1) identically to author
%Control: production of article title (-1) disabled
%Control: page (0) single
%Control: year (1) truncated
%Control: production of eprint (0) enabled
\begin{thebibliography}{80}%
\makeatletter
\providecommand \@ifxundefined [1]{%
 \@ifx{#1\undefined}
}%
\providecommand \@ifnum [1]{%
 \ifnum #1\expandafter \@firstoftwo
 \else \expandafter \@secondoftwo
 \fi
}%
\providecommand \@ifx [1]{%
 \ifx #1\expandafter \@firstoftwo
 \else \expandafter \@secondoftwo
 \fi
}%
\providecommand \natexlab [1]{#1}%
\providecommand \enquote  [1]{``#1''}%
\providecommand \bibnamefont  [1]{#1}%
\providecommand \bibfnamefont [1]{#1}%
\providecommand \citenamefont [1]{#1}%
\providecommand \href@noop [0]{\@secondoftwo}%
\providecommand \href [0]{\begingroup \@sanitize@url \@href}%
\providecommand \@href[1]{\@@startlink{#1}\@@href}%
\providecommand \@@href[1]{\endgroup#1\@@endlink}%
\providecommand \@sanitize@url [0]{\catcode `\\12\catcode `\$12\catcode
  `\&12\catcode `\#12\catcode `\^12\catcode `\_12\catcode `\%12\relax}%
\providecommand \@@startlink[1]{}%
\providecommand \@@endlink[0]{}%
\providecommand \url  [0]{\begingroup\@sanitize@url \@url }%
\providecommand \@url [1]{\endgroup\@href {#1}{\urlprefix }}%
\providecommand \urlprefix  [0]{URL }%
\providecommand \Eprint [0]{\href }%
\providecommand \doibase [0]{http://dx.doi.org/}%
\providecommand \selectlanguage [0]{\@gobble}%
\providecommand \bibinfo  [0]{\@secondoftwo}%
\providecommand \bibfield  [0]{\@secondoftwo}%
\providecommand \translation [1]{[#1]}%
\providecommand \BibitemOpen [0]{}%
\providecommand \bibitemStop [0]{}%
\providecommand \bibitemNoStop [0]{.\EOS\space}%
\providecommand \EOS [0]{\spacefactor3000\relax}%
\providecommand \BibitemShut  [1]{\csname bibitem#1\endcsname}%
\let\auto@bib@innerbib\@empty
%</preamble>
\bibitem [{\citenamefont {Weiler}(1982)}]{Weiler:1982qy}%
  \BibitemOpen
  \bibfield  {author} {\bibinfo {author} {\bibfnamefont {T.~J.}\ \bibnamefont
  {Weiler}},\ }\href {\doibase 10.1103/PhysRevLett.49.234} {\bibfield
  {journal} {\bibinfo  {journal} {Phys. Rev. Lett.}\ }\textbf {\bibinfo
  {volume} {49}},\ \bibinfo {pages} {234} (\bibinfo {year} {1982})}\BibitemShut
  {NoStop}%
\bibitem [{\citenamefont {Argüelles}\ \emph {et~al.}(2019)\citenamefont
  {Argüelles} \emph {et~al.}}]{Arguelles:2019xgp}%
  \BibitemOpen
  \bibfield  {author} {\bibinfo {author} {\bibfnamefont {C.~A.}\ \bibnamefont
  {Argüelles}} \emph {et~al.},\ }\href@noop {} {\  (\bibinfo {year} {2019})},\
  \Eprint {http://arxiv.org/abs/1907.08311} {arXiv:1907.08311 [hep-ph]}
  \BibitemShut {NoStop}%
%%CITATION = ARXIV:1907.08311;%%
\bibitem [{\citenamefont {Tanabashi}\ \emph {et~al.}(2018)\citenamefont
  {Tanabashi} \emph {et~al.}}]{Tanabashi:2018oca}%
  \BibitemOpen
  \bibfield  {author} {\bibinfo {author} {\bibfnamefont {M.}~\bibnamefont
  {Tanabashi}} \emph {et~al.} (\bibinfo {collaboration} {Particle Data
  Group}),\ }\href {\doibase 10.1103/PhysRevD.98.030001} {\bibfield  {journal}
  {\bibinfo  {journal} {Phys. Rev.}\ }\textbf {\bibinfo {volume} {D98}},\
  \bibinfo {pages} {030001} (\bibinfo {year} {2018})}\BibitemShut {NoStop}%
%%CITATION = PHRVA,D98,030001;%%
\bibitem [{\citenamefont {Tzanov}\ \emph {et~al.}(2006)\citenamefont {Tzanov}
  \emph {et~al.}}]{Tzanov:2005kr}%
  \BibitemOpen
  \bibfield  {author} {\bibinfo {author} {\bibfnamefont {M.}~\bibnamefont
  {Tzanov}} \emph {et~al.} (\bibinfo {collaboration} {NuTeV}),\ }\href
  {\doibase 10.1103/PhysRevD.74.012008} {\bibfield  {journal} {\bibinfo
  {journal} {Phys. Rev.}\ }\textbf {\bibinfo {volume} {D74}},\ \bibinfo {pages}
  {012008} (\bibinfo {year} {2006})},\ \Eprint
  {http://arxiv.org/abs/hep-ex/0509010} {arXiv:hep-ex/0509010 [hep-ex]}
  \BibitemShut {NoStop}%
%%CITATION = HEP-EX/0509010;%%
\bibitem [{\citenamefont {Feng}\ \emph {et~al.}(2018)\citenamefont {Feng},
  \citenamefont {Galon}, \citenamefont {Kling},\ and\ \citenamefont
  {Trojanowski}}]{Feng:2017uoz}%
  \BibitemOpen
  \bibfield  {author} {\bibinfo {author} {\bibfnamefont {J.~L.}\ \bibnamefont
  {Feng}}, \bibinfo {author} {\bibfnamefont {I.}~\bibnamefont {Galon}},
  \bibinfo {author} {\bibfnamefont {F.}~\bibnamefont {Kling}}, \ and\ \bibinfo
  {author} {\bibfnamefont {S.}~\bibnamefont {Trojanowski}},\ }\href {\doibase
  10.1103/PhysRevD.97.035001} {\bibfield  {journal} {\bibinfo  {journal} {Phys.
  Rev.}\ }\textbf {\bibinfo {volume} {D97}},\ \bibinfo {pages} {035001}
  (\bibinfo {year} {2018})},\ \Eprint {http://arxiv.org/abs/1708.09389}
  {arXiv:1708.09389 [hep-ph]} \BibitemShut {NoStop}%
%%CITATION = ARXIV:1708.09389;%%
\bibitem [{\citenamefont {Abreu}\ \emph
  {et~al.}(2020{\natexlab{a}})\citenamefont {Abreu} \emph
  {et~al.}}]{Abreu:2019yak}%
  \BibitemOpen
  \bibfield  {author} {\bibinfo {author} {\bibfnamefont {H.}~\bibnamefont
  {Abreu}} \emph {et~al.} (\bibinfo {collaboration} {FASER}),\ }\href {\doibase
  10.1140/epjc/s10052-020-7631-5} {\bibfield  {journal} {\bibinfo  {journal}
  {Eur. Phys. J.}\ }\textbf {\bibinfo {volume} {C80}},\ \bibinfo {pages} {61}
  (\bibinfo {year} {2020}{\natexlab{a}})},\ \Eprint
  {http://arxiv.org/abs/1908.02310} {arXiv:1908.02310 [hep-ex]} \BibitemShut
  {NoStop}%
%%CITATION = ARXIV:1908.02310;%%
\bibitem [{\citenamefont {Abreu}\ \emph
  {et~al.}(2020{\natexlab{b}})\citenamefont {Abreu} \emph
  {et~al.}}]{Abreu:2020ddv}%
  \BibitemOpen
  \bibfield  {author} {\bibinfo {author} {\bibfnamefont {H.}~\bibnamefont
  {Abreu}} \emph {et~al.} (\bibinfo {collaboration} {FASER}),\ }\href@noop {}
  {\  (\bibinfo {year} {2020}{\natexlab{b}})},\ \Eprint
  {http://arxiv.org/abs/2001.03073} {arXiv:2001.03073 [physics.ins-det]}
  \BibitemShut {NoStop}%
%%CITATION = ARXIV:2001.03073;%%
\bibitem [{\citenamefont {Aartsen}\ \emph
  {et~al.}(2017{\natexlab{a}})\citenamefont {Aartsen} \emph
  {et~al.}}]{Aartsen:2017kpd}%
  \BibitemOpen
  \bibfield  {author} {\bibinfo {author} {\bibfnamefont {M.~G.}\ \bibnamefont
  {Aartsen}} \emph {et~al.} (\bibinfo {collaboration} {IceCube}),\ }\href
  {\doibase 10.1038/nature24459} {\bibfield  {journal} {\bibinfo  {journal}
  {Nature}\ }\textbf {\bibinfo {volume} {551}},\ \bibinfo {pages} {596}
  (\bibinfo {year} {2017}{\natexlab{a}})},\ \Eprint
  {http://arxiv.org/abs/1711.08119} {arXiv:1711.08119 [hep-ex]} \BibitemShut
  {NoStop}%
%%CITATION = ARXIV:1711.08119;%%
\bibitem [{\citenamefont {Bustamante}\ and\ \citenamefont
  {Connolly}(2019)}]{Bustamante:2017xuy}%
  \BibitemOpen
  \bibfield  {author} {\bibinfo {author} {\bibfnamefont {M.}~\bibnamefont
  {Bustamante}}\ and\ \bibinfo {author} {\bibfnamefont {A.}~\bibnamefont
  {Connolly}},\ }\href {\doibase 10.1103/PhysRevLett.122.041101} {\bibfield
  {journal} {\bibinfo  {journal} {Phys. Rev. Lett.}\ }\textbf {\bibinfo
  {volume} {122}},\ \bibinfo {pages} {041101} (\bibinfo {year} {2019})},\
  \Eprint {http://arxiv.org/abs/1711.11043} {arXiv:1711.11043 [astro-ph.HE]}
  \BibitemShut {NoStop}%
%%CITATION = ARXIV:1711.11043;%%
\bibitem [{\citenamefont {Anchordoqui}\ \emph {et~al.}(2019)\citenamefont
  {Anchordoqui}, \citenamefont {García~Canal},\ and\ \citenamefont
  {Soriano}}]{Anchordoqui:2019ufu}%
  \BibitemOpen
  \bibfield  {author} {\bibinfo {author} {\bibfnamefont {L.~A.}\ \bibnamefont
  {Anchordoqui}}, \bibinfo {author} {\bibfnamefont {C.}~\bibnamefont
  {García~Canal}}, \ and\ \bibinfo {author} {\bibfnamefont {J.~F.}\
  \bibnamefont {Soriano}},\ }\href {\doibase 10.1103/PhysRevD.100.103001}
  {\bibfield  {journal} {\bibinfo  {journal} {Phys. Rev. D}\ }\textbf {\bibinfo
  {volume} {100}},\ \bibinfo {pages} {103001} (\bibinfo {year} {2019})},\
  \Eprint {http://arxiv.org/abs/1902.10134} {arXiv:1902.10134 [hep-ph]}
  \BibitemShut {NoStop}%
\bibitem [{\citenamefont {Cooper-Sarkar}\ \emph {et~al.}(2011)\citenamefont
  {Cooper-Sarkar}, \citenamefont {Mertsch},\ and\ \citenamefont
  {Sarkar}}]{CooperSarkar:2011pa}%
  \BibitemOpen
  \bibfield  {author} {\bibinfo {author} {\bibfnamefont {A.}~\bibnamefont
  {Cooper-Sarkar}}, \bibinfo {author} {\bibfnamefont {P.}~\bibnamefont
  {Mertsch}}, \ and\ \bibinfo {author} {\bibfnamefont {S.}~\bibnamefont
  {Sarkar}},\ }\href {\doibase 10.1007/JHEP08(2011)042} {\bibfield  {journal}
  {\bibinfo  {journal} {JHEP}\ }\textbf {\bibinfo {volume} {08}},\ \bibinfo
  {pages} {042} (\bibinfo {year} {2011})},\ \Eprint
  {http://arxiv.org/abs/1106.3723} {arXiv:1106.3723 [hep-ph]} \BibitemShut
  {NoStop}%
%%CITATION = ARXIV:1106.3723;%%
\bibitem [{\citenamefont {Connolly}\ \emph {et~al.}(2011)\citenamefont
  {Connolly}, \citenamefont {Thorne},\ and\ \citenamefont
  {Waters}}]{Connolly:2011vc}%
  \BibitemOpen
  \bibfield  {author} {\bibinfo {author} {\bibfnamefont {A.}~\bibnamefont
  {Connolly}}, \bibinfo {author} {\bibfnamefont {R.~S.}\ \bibnamefont
  {Thorne}}, \ and\ \bibinfo {author} {\bibfnamefont {D.}~\bibnamefont
  {Waters}},\ }\href {\doibase 10.1103/PhysRevD.83.113009} {\bibfield
  {journal} {\bibinfo  {journal} {Phys. Rev.}\ }\textbf {\bibinfo {volume}
  {D83}},\ \bibinfo {pages} {113009} (\bibinfo {year} {2011})},\ \Eprint
  {http://arxiv.org/abs/1102.0691} {arXiv:1102.0691 [hep-ph]} \BibitemShut
  {NoStop}%
%%CITATION = ARXIV:1102.0691;%%
\bibitem [{\citenamefont {Argüelles}\ \emph {et~al.}(2015)\citenamefont
  {Argüelles}, \citenamefont {Halzen}, \citenamefont {Wille}, \citenamefont
  {Kroll},\ and\ \citenamefont {Reno}}]{Arguelles:2015wba}%
  \BibitemOpen
  \bibfield  {author} {\bibinfo {author} {\bibfnamefont {C.~A.}\ \bibnamefont
  {Argüelles}}, \bibinfo {author} {\bibfnamefont {F.}~\bibnamefont {Halzen}},
  \bibinfo {author} {\bibfnamefont {L.}~\bibnamefont {Wille}}, \bibinfo
  {author} {\bibfnamefont {M.}~\bibnamefont {Kroll}}, \ and\ \bibinfo {author}
  {\bibfnamefont {M.~H.}\ \bibnamefont {Reno}},\ }\href {\doibase
  10.1103/PhysRevD.92.074040} {\bibfield  {journal} {\bibinfo  {journal} {Phys.
  Rev.}\ }\textbf {\bibinfo {volume} {D92}},\ \bibinfo {pages} {074040}
  (\bibinfo {year} {2015})},\ \Eprint {http://arxiv.org/abs/1504.06639}
  {arXiv:1504.06639 [hep-ph]} \BibitemShut {NoStop}%
%%CITATION = ARXIV:1504.06639;%%
\bibitem [{\citenamefont {Garcia}\ \emph {et~al.}(2020)\citenamefont {Garcia},
  \citenamefont {Gauld}, \citenamefont {Heijboer},\ and\ \citenamefont
  {Rojo}}]{Garcia:2020jwr}%
  \BibitemOpen
  \bibfield  {author} {\bibinfo {author} {\bibfnamefont {A.}~\bibnamefont
  {Garcia}}, \bibinfo {author} {\bibfnamefont {R.}~\bibnamefont {Gauld}},
  \bibinfo {author} {\bibfnamefont {A.}~\bibnamefont {Heijboer}}, \ and\
  \bibinfo {author} {\bibfnamefont {J.}~\bibnamefont {Rojo}},\ }\href@noop {}
  {\  (\bibinfo {year} {2020})},\ \Eprint {http://arxiv.org/abs/2004.04756}
  {arXiv:2004.04756 [hep-ph]} \BibitemShut {NoStop}%
\bibitem [{\citenamefont {Bertone}\ \emph {et~al.}(2019)\citenamefont
  {Bertone}, \citenamefont {Gauld},\ and\ \citenamefont
  {Rojo}}]{Bertone:2018dse}%
  \BibitemOpen
  \bibfield  {author} {\bibinfo {author} {\bibfnamefont {V.}~\bibnamefont
  {Bertone}}, \bibinfo {author} {\bibfnamefont {R.}~\bibnamefont {Gauld}}, \
  and\ \bibinfo {author} {\bibfnamefont {J.}~\bibnamefont {Rojo}},\ }\href
  {\doibase 10.1007/JHEP01(2019)217} {\bibfield  {journal} {\bibinfo  {journal}
  {JHEP}\ }\textbf {\bibinfo {volume} {01}},\ \bibinfo {pages} {217} (\bibinfo
  {year} {2019})},\ \Eprint {http://arxiv.org/abs/1808.02034} {arXiv:1808.02034
  [hep-ph]} \BibitemShut {NoStop}%
\bibitem [{\citenamefont {Chen}\ \emph {et~al.}(2014)\citenamefont {Chen},
  \citenamefont {Bhupal~Dev},\ and\ \citenamefont {Soni}}]{Chen:2013dza}%
  \BibitemOpen
  \bibfield  {author} {\bibinfo {author} {\bibfnamefont {C.-Y.}\ \bibnamefont
  {Chen}}, \bibinfo {author} {\bibfnamefont {P.}~\bibnamefont {Bhupal~Dev}}, \
  and\ \bibinfo {author} {\bibfnamefont {A.}~\bibnamefont {Soni}},\ }\href
  {\doibase 10.1103/PhysRevD.89.033012} {\bibfield  {journal} {\bibinfo
  {journal} {Phys. Rev. D}\ }\textbf {\bibinfo {volume} {89}},\ \bibinfo
  {pages} {033012} (\bibinfo {year} {2014})},\ \Eprint
  {http://arxiv.org/abs/1309.1764} {arXiv:1309.1764 [hep-ph]} \BibitemShut
  {NoStop}%
\bibitem [{\citenamefont {Hobbs}\ \emph {et~al.}(2019)\citenamefont {Hobbs},
  \citenamefont {Wang}, \citenamefont {Nadolsky},\ and\ \citenamefont
  {Olness}}]{Hobbs:2019sut}%
  \BibitemOpen
  \bibfield  {author} {\bibinfo {author} {\bibfnamefont {T.}~\bibnamefont
  {Hobbs}}, \bibinfo {author} {\bibfnamefont {B.-T.}\ \bibnamefont {Wang}},
  \bibinfo {author} {\bibfnamefont {P.~M.}\ \bibnamefont {Nadolsky}}, \ and\
  \bibinfo {author} {\bibfnamefont {F.~I.}\ \bibnamefont {Olness}},\ }\href
  {\doibase 10.22323/1.352.0247} {\bibfield  {journal} {\bibinfo  {journal}
  {PoS}\ }\textbf {\bibinfo {volume} {DIS2019}},\ \bibinfo {pages} {247}
  (\bibinfo {year} {2019})},\ \Eprint {http://arxiv.org/abs/1907.00988}
  {arXiv:1907.00988 [hep-ph]} \BibitemShut {NoStop}%
\bibitem [{\citenamefont {Alvarez-Muniz}\ \emph {et~al.}(2002)\citenamefont
  {Alvarez-Muniz}, \citenamefont {Halzen}, \citenamefont {Han},\ and\
  \citenamefont {Hooper}}]{AlvarezMuniz:2001mk}%
  \BibitemOpen
  \bibfield  {author} {\bibinfo {author} {\bibfnamefont {J.}~\bibnamefont
  {Alvarez-Muniz}}, \bibinfo {author} {\bibfnamefont {F.}~\bibnamefont
  {Halzen}}, \bibinfo {author} {\bibfnamefont {T.}~\bibnamefont {Han}}, \ and\
  \bibinfo {author} {\bibfnamefont {D.}~\bibnamefont {Hooper}},\ }\href
  {\doibase 10.1103/PhysRevLett.88.021301} {\bibfield  {journal} {\bibinfo
  {journal} {Phys. Rev. Lett.}\ }\textbf {\bibinfo {volume} {88}},\ \bibinfo
  {pages} {021301} (\bibinfo {year} {2002})},\ \Eprint
  {http://arxiv.org/abs/hep-ph/0107057} {arXiv:hep-ph/0107057} \BibitemShut
  {NoStop}%
\bibitem [{\citenamefont {Ellis}\ \emph {et~al.}(2016)\citenamefont {Ellis},
  \citenamefont {Sakurai},\ and\ \citenamefont {Spannowsky}}]{Ellis:2016dgb}%
  \BibitemOpen
  \bibfield  {author} {\bibinfo {author} {\bibfnamefont {J.}~\bibnamefont
  {Ellis}}, \bibinfo {author} {\bibfnamefont {K.}~\bibnamefont {Sakurai}}, \
  and\ \bibinfo {author} {\bibfnamefont {M.}~\bibnamefont {Spannowsky}},\
  }\href {\doibase 10.1007/JHEP05(2016)085} {\bibfield  {journal} {\bibinfo
  {journal} {JHEP}\ }\textbf {\bibinfo {volume} {05}},\ \bibinfo {pages} {085}
  (\bibinfo {year} {2016})},\ \Eprint {http://arxiv.org/abs/1603.06573}
  {arXiv:1603.06573 [hep-ph]} \BibitemShut {NoStop}%
\bibitem [{\citenamefont {Henley}\ and\ \citenamefont
  {Jalilian-Marian}(2006)}]{Henley:2005ms}%
  \BibitemOpen
  \bibfield  {author} {\bibinfo {author} {\bibfnamefont {E.~M.}\ \bibnamefont
  {Henley}}\ and\ \bibinfo {author} {\bibfnamefont {J.}~\bibnamefont
  {Jalilian-Marian}},\ }\href {\doibase 10.1103/PhysRevD.73.094004} {\bibfield
  {journal} {\bibinfo  {journal} {Phys. Rev. D}\ }\textbf {\bibinfo {volume}
  {73}},\ \bibinfo {pages} {094004} (\bibinfo {year} {2006})},\ \Eprint
  {http://arxiv.org/abs/hep-ph/0512220} {arXiv:hep-ph/0512220} \BibitemShut
  {NoStop}%
\bibitem [{\citenamefont {Greisen}(1966)}]{Greisen:1966jv}%
  \BibitemOpen
  \bibfield  {author} {\bibinfo {author} {\bibfnamefont {K.}~\bibnamefont
  {Greisen}},\ }\href {\doibase 10.1103/PhysRevLett.16.748} {\bibfield
  {journal} {\bibinfo  {journal} {Phys. Rev. Lett.}\ }\textbf {\bibinfo
  {volume} {16}},\ \bibinfo {pages} {748} (\bibinfo {year} {1966})}\BibitemShut
  {NoStop}%
\bibitem [{\citenamefont {Zatsepin}\ and\ \citenamefont
  {Kuzmin}(1966)}]{Zatsepin:1966jv}%
  \BibitemOpen
  \bibfield  {author} {\bibinfo {author} {\bibfnamefont {G.}~\bibnamefont
  {Zatsepin}}\ and\ \bibinfo {author} {\bibfnamefont {V.}~\bibnamefont
  {Kuzmin}},\ }\href@noop {} {\bibfield  {journal} {\bibinfo  {journal} {JETP
  Lett.}\ }\textbf {\bibinfo {volume} {4}},\ \bibinfo {pages} {78} (\bibinfo
  {year} {1966})}\BibitemShut {NoStop}%
\bibitem [{\citenamefont {Stecker}\ \emph {et~al.}(1991)\citenamefont
  {Stecker}, \citenamefont {Done}, \citenamefont {Salamon},\ and\ \citenamefont
  {Sommers}}]{Stecker:1991vm}%
  \BibitemOpen
  \bibfield  {author} {\bibinfo {author} {\bibfnamefont {F.}~\bibnamefont
  {Stecker}}, \bibinfo {author} {\bibfnamefont {C.}~\bibnamefont {Done}},
  \bibinfo {author} {\bibfnamefont {M.}~\bibnamefont {Salamon}}, \ and\
  \bibinfo {author} {\bibfnamefont {P.}~\bibnamefont {Sommers}},\ }\href
  {\doibase 10.1103/PhysRevLett.66.2697} {\bibfield  {journal} {\bibinfo
  {journal} {Phys. Rev. Lett.}\ }\textbf {\bibinfo {volume} {66}},\ \bibinfo
  {pages} {2697} (\bibinfo {year} {1991})},\ \bibinfo {note} {[Erratum:
  Phys.Rev.Lett. 69, 2738 (1992)]}\BibitemShut {NoStop}%
\bibitem [{\citenamefont {Waxman}\ and\ \citenamefont
  {Bahcall}(1999)}]{Waxman:1998yy}%
  \BibitemOpen
  \bibfield  {author} {\bibinfo {author} {\bibfnamefont {E.}~\bibnamefont
  {Waxman}}\ and\ \bibinfo {author} {\bibfnamefont {J.~N.}\ \bibnamefont
  {Bahcall}},\ }\href {\doibase 10.1103/PhysRevD.59.023002} {\bibfield
  {journal} {\bibinfo  {journal} {Phys. Rev. D}\ }\textbf {\bibinfo {volume}
  {59}},\ \bibinfo {pages} {023002} (\bibinfo {year} {1999})},\ \Eprint
  {http://arxiv.org/abs/hep-ph/9807282} {arXiv:hep-ph/9807282} \BibitemShut
  {NoStop}%
\bibitem [{\citenamefont {Murase}\ \emph {et~al.}(2006)\citenamefont {Murase},
  \citenamefont {Ioka}, \citenamefont {Nagataki},\ and\ \citenamefont
  {Nakamura}}]{Murase:2006mm}%
  \BibitemOpen
  \bibfield  {author} {\bibinfo {author} {\bibfnamefont {K.}~\bibnamefont
  {Murase}}, \bibinfo {author} {\bibfnamefont {K.}~\bibnamefont {Ioka}},
  \bibinfo {author} {\bibfnamefont {S.}~\bibnamefont {Nagataki}}, \ and\
  \bibinfo {author} {\bibfnamefont {T.}~\bibnamefont {Nakamura}},\ }\href
  {\doibase 10.1086/509323} {\bibfield  {journal} {\bibinfo  {journal}
  {Astrophys. J. Lett.}\ }\textbf {\bibinfo {volume} {651}},\ \bibinfo {pages}
  {L5} (\bibinfo {year} {2006})},\ \Eprint
  {http://arxiv.org/abs/astro-ph/0607104} {arXiv:astro-ph/0607104} \BibitemShut
  {NoStop}%
\bibitem [{\citenamefont {Murase}(2007)}]{Murase:2007yt}%
  \BibitemOpen
  \bibfield  {author} {\bibinfo {author} {\bibfnamefont {K.}~\bibnamefont
  {Murase}},\ }\href {\doibase 10.1103/PhysRevD.76.123001} {\bibfield
  {journal} {\bibinfo  {journal} {Phys. Rev.}\ }\textbf {\bibinfo {volume}
  {D76}},\ \bibinfo {pages} {123001} (\bibinfo {year} {2007})},\ \Eprint
  {http://arxiv.org/abs/0707.1140} {arXiv:0707.1140 [astro-ph]} \BibitemShut
  {NoStop}%
%%CITATION = ARXIV:0707.1140;%%
\bibitem [{\citenamefont {Murase}\ \emph
  {et~al.}(2008{\natexlab{a}})\citenamefont {Murase}, \citenamefont {Ioka},
  \citenamefont {Nagataki},\ and\ \citenamefont {Nakamura}}]{Murase:2008mr}%
  \BibitemOpen
  \bibfield  {author} {\bibinfo {author} {\bibfnamefont {K.}~\bibnamefont
  {Murase}}, \bibinfo {author} {\bibfnamefont {K.}~\bibnamefont {Ioka}},
  \bibinfo {author} {\bibfnamefont {S.}~\bibnamefont {Nagataki}}, \ and\
  \bibinfo {author} {\bibfnamefont {T.}~\bibnamefont {Nakamura}},\ }\href
  {\doibase 10.1103/PhysRevD.78.023005} {\bibfield  {journal} {\bibinfo
  {journal} {Phys. Rev. D}\ }\textbf {\bibinfo {volume} {78}},\ \bibinfo
  {pages} {023005} (\bibinfo {year} {2008}{\natexlab{a}})},\ \Eprint
  {http://arxiv.org/abs/0801.2861} {arXiv:0801.2861 [astro-ph]} \BibitemShut
  {NoStop}%
\bibitem [{\citenamefont {Murase}\ \emph
  {et~al.}(2008{\natexlab{b}})\citenamefont {Murase}, \citenamefont {Inoue},\
  and\ \citenamefont {Nagataki}}]{Murase:2008yt}%
  \BibitemOpen
  \bibfield  {author} {\bibinfo {author} {\bibfnamefont {K.}~\bibnamefont
  {Murase}}, \bibinfo {author} {\bibfnamefont {S.}~\bibnamefont {Inoue}}, \
  and\ \bibinfo {author} {\bibfnamefont {S.}~\bibnamefont {Nagataki}},\ }\href
  {\doibase 10.1086/595882} {\bibfield  {journal} {\bibinfo  {journal}
  {Astrophys. J.}\ }\textbf {\bibinfo {volume} {689}},\ \bibinfo {pages} {L105}
  (\bibinfo {year} {2008}{\natexlab{b}})},\ \Eprint
  {http://arxiv.org/abs/0805.0104} {arXiv:0805.0104 [astro-ph]} \BibitemShut
  {NoStop}%
%%CITATION = ARXIV:0805.0104;%%
\bibitem [{\citenamefont {Kotera}\ \emph {et~al.}(2009)\citenamefont {Kotera},
  \citenamefont {Allard}, \citenamefont {Murase}, \citenamefont {Aoi},
  \citenamefont {Dubois}, \citenamefont {Pierog},\ and\ \citenamefont
  {Nagataki}}]{Kotera:2009ms}%
  \BibitemOpen
  \bibfield  {author} {\bibinfo {author} {\bibfnamefont {K.}~\bibnamefont
  {Kotera}}, \bibinfo {author} {\bibfnamefont {D.}~\bibnamefont {Allard}},
  \bibinfo {author} {\bibfnamefont {K.}~\bibnamefont {Murase}}, \bibinfo
  {author} {\bibfnamefont {J.}~\bibnamefont {Aoi}}, \bibinfo {author}
  {\bibfnamefont {Y.}~\bibnamefont {Dubois}}, \bibinfo {author} {\bibfnamefont
  {T.}~\bibnamefont {Pierog}}, \ and\ \bibinfo {author} {\bibfnamefont
  {S.}~\bibnamefont {Nagataki}},\ }\href {\doibase 10.1088/0004-637X/707/1/370}
  {\bibfield  {journal} {\bibinfo  {journal} {Astrophys. J.}\ }\textbf
  {\bibinfo {volume} {707}},\ \bibinfo {pages} {370} (\bibinfo {year}
  {2009})},\ \Eprint {http://arxiv.org/abs/0907.2433} {arXiv:0907.2433
  [astro-ph.HE]} \BibitemShut {NoStop}%
\bibitem [{\citenamefont {Murase}\ \emph {et~al.}(2009)\citenamefont {Murase},
  \citenamefont {Meszaros},\ and\ \citenamefont {Zhang}}]{Murase:2009pg}%
  \BibitemOpen
  \bibfield  {author} {\bibinfo {author} {\bibfnamefont {K.}~\bibnamefont
  {Murase}}, \bibinfo {author} {\bibfnamefont {P.}~\bibnamefont {Meszaros}}, \
  and\ \bibinfo {author} {\bibfnamefont {B.}~\bibnamefont {Zhang}},\ }\href
  {\doibase 10.1103/PhysRevD.79.103001} {\bibfield  {journal} {\bibinfo
  {journal} {Phys. Rev. D}\ }\textbf {\bibinfo {volume} {79}},\ \bibinfo
  {pages} {103001} (\bibinfo {year} {2009})},\ \Eprint
  {http://arxiv.org/abs/0904.2509} {arXiv:0904.2509 [astro-ph.HE]} \BibitemShut
  {NoStop}%
\bibitem [{\citenamefont {He}\ \emph {et~al.}(2012)\citenamefont {He},
  \citenamefont {Liu}, \citenamefont {Wang}, \citenamefont {Nagataki},
  \citenamefont {Murase},\ and\ \citenamefont {Dai}}]{He:2012tq}%
  \BibitemOpen
  \bibfield  {author} {\bibinfo {author} {\bibfnamefont {H.-N.}\ \bibnamefont
  {He}}, \bibinfo {author} {\bibfnamefont {R.-Y.}\ \bibnamefont {Liu}},
  \bibinfo {author} {\bibfnamefont {X.-Y.}\ \bibnamefont {Wang}}, \bibinfo
  {author} {\bibfnamefont {S.}~\bibnamefont {Nagataki}}, \bibinfo {author}
  {\bibfnamefont {K.}~\bibnamefont {Murase}}, \ and\ \bibinfo {author}
  {\bibfnamefont {Z.-G.}\ \bibnamefont {Dai}},\ }\href {\doibase
  10.1088/0004-637X/752/1/29} {\bibfield  {journal} {\bibinfo  {journal}
  {Astrophys. J.}\ }\textbf {\bibinfo {volume} {752}},\ \bibinfo {pages} {29}
  (\bibinfo {year} {2012})},\ \Eprint {http://arxiv.org/abs/1204.0857}
  {arXiv:1204.0857 [astro-ph.HE]} \BibitemShut {NoStop}%
\bibitem [{\citenamefont {Fang}\ \emph {et~al.}(2014)\citenamefont {Fang},
  \citenamefont {Kotera}, \citenamefont {Murase},\ and\ \citenamefont
  {Olinto}}]{Fang:2013vla}%
  \BibitemOpen
  \bibfield  {author} {\bibinfo {author} {\bibfnamefont {K.}~\bibnamefont
  {Fang}}, \bibinfo {author} {\bibfnamefont {K.}~\bibnamefont {Kotera}},
  \bibinfo {author} {\bibfnamefont {K.}~\bibnamefont {Murase}}, \ and\ \bibinfo
  {author} {\bibfnamefont {A.~V.}\ \bibnamefont {Olinto}},\ }\href {\doibase
  10.1103/PhysRevD.90.103005, 10.1103/PhysRevD.92.129901} {\bibfield  {journal}
  {\bibinfo  {journal} {Phys. Rev.}\ }\textbf {\bibinfo {volume} {D90}},\
  \bibinfo {pages} {103005} (\bibinfo {year} {2014})},\ \bibinfo {note}
  {[Erratum: Phys. Rev.D92,no.12,129901(2015); Phys. Rev.D90,103005(2014)]},\
  \Eprint {http://arxiv.org/abs/1311.2044} {arXiv:1311.2044 [astro-ph.HE]}
  \BibitemShut {NoStop}%
%%CITATION = ARXIV:1311.2044;%%
\bibitem [{\citenamefont {Murase}\ \emph {et~al.}(2014)\citenamefont {Murase},
  \citenamefont {Inoue},\ and\ \citenamefont {Dermer}}]{Murase:2014foa}%
  \BibitemOpen
  \bibfield  {author} {\bibinfo {author} {\bibfnamefont {K.}~\bibnamefont
  {Murase}}, \bibinfo {author} {\bibfnamefont {Y.}~\bibnamefont {Inoue}}, \
  and\ \bibinfo {author} {\bibfnamefont {C.~D.}\ \bibnamefont {Dermer}},\
  }\href {\doibase 10.1103/PhysRevD.90.023007} {\bibfield  {journal} {\bibinfo
  {journal} {Phys. Rev. D}\ }\textbf {\bibinfo {volume} {90}},\ \bibinfo
  {pages} {023007} (\bibinfo {year} {2014})},\ \Eprint
  {http://arxiv.org/abs/1403.4089} {arXiv:1403.4089 [astro-ph.HE]} \BibitemShut
  {NoStop}%
\bibitem [{\citenamefont {Bustamante}\ \emph {et~al.}(2015)\citenamefont
  {Bustamante}, \citenamefont {Baerwald}, \citenamefont {Murase},\ and\
  \citenamefont {Winter}}]{Bustamante:2014oka}%
  \BibitemOpen
  \bibfield  {author} {\bibinfo {author} {\bibfnamefont {M.}~\bibnamefont
  {Bustamante}}, \bibinfo {author} {\bibfnamefont {P.}~\bibnamefont
  {Baerwald}}, \bibinfo {author} {\bibfnamefont {K.}~\bibnamefont {Murase}}, \
  and\ \bibinfo {author} {\bibfnamefont {W.}~\bibnamefont {Winter}},\ }\href
  {\doibase 10.1038/ncomms7783} {\bibfield  {journal} {\bibinfo  {journal}
  {Nature Commun.}\ }\textbf {\bibinfo {volume} {6}},\ \bibinfo {pages} {6783}
  (\bibinfo {year} {2015})},\ \Eprint {http://arxiv.org/abs/1409.2874}
  {arXiv:1409.2874 [astro-ph.HE]} \BibitemShut {NoStop}%
\bibitem [{\citenamefont {Murase}(2017)}]{Murase:2015ndr}%
  \BibitemOpen
  \bibfield  {author} {\bibinfo {author} {\bibfnamefont {K.}~\bibnamefont
  {Murase}},\ }in\ \href {\doibase 10.1142/9789814759410_0002} {\emph {\bibinfo
  {booktitle} {Neutrino Astronomy: Current Status, Future Prospects}}},\
  \bibinfo {editor} {edited by\ \bibinfo {editor} {\bibfnamefont
  {T.}~\bibnamefont {Gaisser}}\ and\ \bibinfo {editor} {\bibfnamefont
  {A.}~\bibnamefont {Karle}}}\ (\bibinfo {year} {2017})\ pp.\ \bibinfo {pages}
  {15--31},\ \Eprint {http://arxiv.org/abs/1511.01590} {arXiv:1511.01590
  [astro-ph.HE]} \BibitemShut {NoStop}%
%%CITATION = ARXIV:1511.01590;%%
\bibitem [{\citenamefont {Bustamante}\ \emph {et~al.}(2017)\citenamefont
  {Bustamante}, \citenamefont {Murase}, \citenamefont {Winter},\ and\
  \citenamefont {Heinze}}]{Bustamante:2016wpu}%
  \BibitemOpen
  \bibfield  {author} {\bibinfo {author} {\bibfnamefont {M.}~\bibnamefont
  {Bustamante}}, \bibinfo {author} {\bibfnamefont {K.}~\bibnamefont {Murase}},
  \bibinfo {author} {\bibfnamefont {W.}~\bibnamefont {Winter}}, \ and\ \bibinfo
  {author} {\bibfnamefont {J.}~\bibnamefont {Heinze}},\ }\href {\doibase
  10.3847/1538-4357/837/1/33} {\bibfield  {journal} {\bibinfo  {journal}
  {Astrophys. J.}\ }\textbf {\bibinfo {volume} {837}},\ \bibinfo {pages} {33}
  (\bibinfo {year} {2017})},\ \Eprint {http://arxiv.org/abs/1606.02325}
  {arXiv:1606.02325 [astro-ph.HE]} \BibitemShut {NoStop}%
\bibitem [{\citenamefont {Denton}\ and\ \citenamefont
  {Tamborra}(2018)}]{Denton:2017jwk}%
  \BibitemOpen
  \bibfield  {author} {\bibinfo {author} {\bibfnamefont {P.~B.}\ \bibnamefont
  {Denton}}\ and\ \bibinfo {author} {\bibfnamefont {I.}~\bibnamefont
  {Tamborra}},\ }\href {\doibase 10.3847/1538-4357/aaab4a} {\bibfield
  {journal} {\bibinfo  {journal} {Astrophys. J.}\ }\textbf {\bibinfo {volume}
  {855}},\ \bibinfo {pages} {37} (\bibinfo {year} {2018})},\ \Eprint
  {http://arxiv.org/abs/1711.00470} {arXiv:1711.00470 [astro-ph.HE]}
  \BibitemShut {NoStop}%
%%CITATION = ARXIV:1711.00470;%%
\bibitem [{\citenamefont {Fang}\ and\ \citenamefont
  {Murase}(2018)}]{Fang:2017zjf}%
  \BibitemOpen
  \bibfield  {author} {\bibinfo {author} {\bibfnamefont {K.}~\bibnamefont
  {Fang}}\ and\ \bibinfo {author} {\bibfnamefont {K.}~\bibnamefont {Murase}},\
  }\href {\doibase 10.1038/s41567-017-0025-4} {\bibfield  {journal} {\bibinfo
  {journal} {Nature Phys.}\ }\textbf {\bibinfo {volume} {14}},\ \bibinfo
  {pages} {396} (\bibinfo {year} {2018})},\ \Eprint
  {http://arxiv.org/abs/1704.00015} {arXiv:1704.00015 [astro-ph.HE]}
  \BibitemShut {NoStop}%
%%CITATION = ARXIV:1704.00015;%%
\bibitem [{\citenamefont {Rodrigues}\ \emph {et~al.}(2020)\citenamefont
  {Rodrigues}, \citenamefont {Heinze}, \citenamefont {Palladino}, \citenamefont
  {van Vliet},\ and\ \citenamefont {Winter}}]{Rodrigues:2020pli}%
  \BibitemOpen
  \bibfield  {author} {\bibinfo {author} {\bibfnamefont {X.}~\bibnamefont
  {Rodrigues}}, \bibinfo {author} {\bibfnamefont {J.}~\bibnamefont {Heinze}},
  \bibinfo {author} {\bibfnamefont {A.}~\bibnamefont {Palladino}}, \bibinfo
  {author} {\bibfnamefont {A.}~\bibnamefont {van Vliet}}, \ and\ \bibinfo
  {author} {\bibfnamefont {W.}~\bibnamefont {Winter}},\ }\href@noop {} {\
  (\bibinfo {year} {2020})},\ \Eprint {http://arxiv.org/abs/2003.08392}
  {arXiv:2003.08392 [astro-ph.HE]} \BibitemShut {NoStop}%
\bibitem [{\citenamefont {Righi}\ \emph {et~al.}(2020)\citenamefont {Righi},
  \citenamefont {Palladino}, \citenamefont {Tavecchio},\ and\ \citenamefont
  {Vissani}}]{Righi:2020ufi}%
  \BibitemOpen
  \bibfield  {author} {\bibinfo {author} {\bibfnamefont {C.}~\bibnamefont
  {Righi}}, \bibinfo {author} {\bibfnamefont {A.}~\bibnamefont {Palladino}},
  \bibinfo {author} {\bibfnamefont {F.}~\bibnamefont {Tavecchio}}, \ and\
  \bibinfo {author} {\bibfnamefont {F.}~\bibnamefont {Vissani}},\ }\href@noop
  {} {\  (\bibinfo {year} {2020})},\ \Eprint {http://arxiv.org/abs/2003.08701}
  {arXiv:2003.08701 [astro-ph.HE]} \BibitemShut {NoStop}%
\bibitem [{\citenamefont {Aartsen}\ \emph {et~al.}(2013)\citenamefont {Aartsen}
  \emph {et~al.}}]{Aartsen:2013jdh}%
  \BibitemOpen
  \bibfield  {author} {\bibinfo {author} {\bibfnamefont {M.~G.}\ \bibnamefont
  {Aartsen}} \emph {et~al.} (\bibinfo {collaboration} {IceCube}),\ }\href
  {\doibase 10.1126/science.1242856} {\bibfield  {journal} {\bibinfo  {journal}
  {Science}\ }\textbf {\bibinfo {volume} {342}},\ \bibinfo {pages} {1242856}
  (\bibinfo {year} {2013})},\ \Eprint {http://arxiv.org/abs/1311.5238}
  {arXiv:1311.5238 [astro-ph.HE]} \BibitemShut {NoStop}%
%%CITATION = ARXIV:1311.5238;%%
\bibitem [{\citenamefont {Barwick}\ \emph {et~al.}(2006)\citenamefont {Barwick}
  \emph {et~al.}}]{Barwick:2005hn}%
  \BibitemOpen
  \bibfield  {author} {\bibinfo {author} {\bibfnamefont {S.~W.}\ \bibnamefont
  {Barwick}} \emph {et~al.} (\bibinfo {collaboration} {ANITA}),\ }\href
  {\doibase 10.1103/PhysRevLett.96.171101} {\bibfield  {journal} {\bibinfo
  {journal} {Phys. Rev. Lett.}\ }\textbf {\bibinfo {volume} {96}},\ \bibinfo
  {pages} {171101} (\bibinfo {year} {2006})},\ \Eprint
  {http://arxiv.org/abs/astro-ph/0512265} {arXiv:astro-ph/0512265 [astro-ph]}
  \BibitemShut {NoStop}%
%%CITATION = ASTRO-PH/0512265;%%
\bibitem [{\citenamefont {Aartsen}\ \emph {et~al.}(2016)\citenamefont {Aartsen}
  \emph {et~al.}}]{Aartsen:2016ngq}%
  \BibitemOpen
  \bibfield  {author} {\bibinfo {author} {\bibfnamefont {M.~G.}\ \bibnamefont
  {Aartsen}} \emph {et~al.} (\bibinfo {collaboration} {IceCube}),\ }\href
  {\doibase 10.1103/PhysRevLett.117.241101, 10.1103/PhysRevLett.119.259902}
  {\bibfield  {journal} {\bibinfo  {journal} {Phys. Rev. Lett.}\ }\textbf
  {\bibinfo {volume} {117}},\ \bibinfo {pages} {241101} (\bibinfo {year}
  {2016})},\ \bibinfo {note} {[Erratum: Phys. Rev.
  Lett.119,no.25,259902(2017)]},\ \Eprint {http://arxiv.org/abs/1607.05886}
  {arXiv:1607.05886 [astro-ph.HE]} \BibitemShut {NoStop}%
%%CITATION = ARXIV:1607.05886;%%
\bibitem [{\citenamefont {Aab}\ \emph {et~al.}(2015)\citenamefont {Aab} \emph
  {et~al.}}]{Aab:2015kma}%
  \BibitemOpen
  \bibfield  {author} {\bibinfo {author} {\bibfnamefont {A.}~\bibnamefont
  {Aab}} \emph {et~al.} (\bibinfo {collaboration} {Pierre Auger}),\ }\href
  {\doibase 10.1103/PhysRevD.91.092008} {\bibfield  {journal} {\bibinfo
  {journal} {Phys. Rev.}\ }\textbf {\bibinfo {volume} {D91}},\ \bibinfo {pages}
  {092008} (\bibinfo {year} {2015})},\ \Eprint
  {http://arxiv.org/abs/1504.05397} {arXiv:1504.05397 [astro-ph.HE]}
  \BibitemShut {NoStop}%
%%CITATION = ARXIV:1504.05397;%%
\bibitem [{\citenamefont {Álvarez Muñiz}\ \emph {et~al.}(2020)\citenamefont
  {Álvarez Muñiz} \emph {et~al.}}]{Alvarez-Muniz:2018bhp}%
  \BibitemOpen
  \bibfield  {author} {\bibinfo {author} {\bibfnamefont {J.}~\bibnamefont
  {Álvarez Muñiz}} \emph {et~al.} (\bibinfo {collaboration} {GRAND}),\ }\href
  {\doibase 10.1007/s11433-018-9385-7} {\bibfield  {journal} {\bibinfo
  {journal} {Sci. China Phys. Mech. Astron.}\ }\textbf {\bibinfo {volume}
  {63}},\ \bibinfo {pages} {219501} (\bibinfo {year} {2020})},\ \Eprint
  {http://arxiv.org/abs/1810.09994} {arXiv:1810.09994 [astro-ph.HE]}
  \BibitemShut {NoStop}%
%%CITATION = ARXIV:1810.09994;%%
\bibitem [{\citenamefont {Olinto}\ \emph {et~al.}(2018)\citenamefont {Olinto}
  \emph {et~al.}}]{Olinto:2017xbi}%
  \BibitemOpen
  \bibfield  {author} {\bibinfo {author} {\bibfnamefont {A.~V.}\ \bibnamefont
  {Olinto}} \emph {et~al.},\ }\bibfield  {booktitle} {\emph {\bibinfo
  {booktitle} {{The Fluorescence detector Array of Single-pixel Telescopes:
  Contributions to the 35th International Cosmic Ray Conference (ICRC
  2017)}}},\ }\href {\doibase 10.22323/1.301.0542} {\bibfield  {journal}
  {\bibinfo  {journal} {PoS}\ }\textbf {\bibinfo {volume} {ICRC2017}},\
  \bibinfo {pages} {542} (\bibinfo {year} {2018})},\ \bibinfo {note}
  {[35,542(2017)]},\ \Eprint {http://arxiv.org/abs/1708.07599}
  {arXiv:1708.07599 [astro-ph.IM]} \BibitemShut {NoStop}%
%%CITATION = ARXIV:1708.07599;%%
\bibitem [{\citenamefont {Alves~Batista}\ \emph {et~al.}(2019)\citenamefont
  {Alves~Batista}, \citenamefont {de~Almeida}, \citenamefont {Lago},\ and\
  \citenamefont {Kotera}}]{AlvesBatista:2018zui}%
  \BibitemOpen
  \bibfield  {author} {\bibinfo {author} {\bibfnamefont {R.}~\bibnamefont
  {Alves~Batista}}, \bibinfo {author} {\bibfnamefont {R.~M.}\ \bibnamefont
  {de~Almeida}}, \bibinfo {author} {\bibfnamefont {B.}~\bibnamefont {Lago}}, \
  and\ \bibinfo {author} {\bibfnamefont {K.}~\bibnamefont {Kotera}},\ }\href
  {\doibase 10.1088/1475-7516/2019/01/002} {\bibfield  {journal} {\bibinfo
  {journal} {JCAP}\ }\textbf {\bibinfo {volume} {1901}},\ \bibinfo {pages}
  {002} (\bibinfo {year} {2019})},\ \Eprint {http://arxiv.org/abs/1806.10879}
  {arXiv:1806.10879 [astro-ph.HE]} \BibitemShut {NoStop}%
%%CITATION = ARXIV:1806.10879;%%
\bibitem [{\citenamefont {Møller}\ \emph {et~al.}(2019)\citenamefont
  {Møller}, \citenamefont {Denton},\ and\ \citenamefont
  {Tamborra}}]{Moller:2018isk}%
  \BibitemOpen
  \bibfield  {author} {\bibinfo {author} {\bibfnamefont {K.}~\bibnamefont
  {Møller}}, \bibinfo {author} {\bibfnamefont {P.~B.}\ \bibnamefont {Denton}},
  \ and\ \bibinfo {author} {\bibfnamefont {I.}~\bibnamefont {Tamborra}},\
  }\href {\doibase 10.1088/1475-7516/2019/05/047} {\bibfield  {journal}
  {\bibinfo  {journal} {JCAP}\ }\textbf {\bibinfo {volume} {05}},\ \bibinfo
  {pages} {047} (\bibinfo {year} {2019})},\ \Eprint
  {http://arxiv.org/abs/1809.04866} {arXiv:1809.04866 [astro-ph.HE]}
  \BibitemShut {NoStop}%
\bibitem [{\citenamefont {Aartsen}\ \emph {et~al.}(2014)\citenamefont {Aartsen}
  \emph {et~al.}}]{Aartsen:2014njl}%
  \BibitemOpen
  \bibfield  {author} {\bibinfo {author} {\bibfnamefont {M.}~\bibnamefont
  {Aartsen}} \emph {et~al.} (\bibinfo {collaboration} {IceCube}),\ }\href@noop
  {} {\  (\bibinfo {year} {2014})},\ \Eprint {http://arxiv.org/abs/1412.5106}
  {arXiv:1412.5106 [astro-ph.HE]} \BibitemShut {NoStop}%
\bibitem [{\citenamefont {Romero-Wolf}\ \emph {et~al.}(2015)\citenamefont
  {Romero-Wolf} \emph {et~al.}}]{Romero-Wolf:2014pua}%
  \BibitemOpen
  \bibfield  {author} {\bibinfo {author} {\bibfnamefont {A.}~\bibnamefont
  {Romero-Wolf}} \emph {et~al.},\ }\href {\doibase
  10.1016/j.astropartphys.2014.06.006} {\bibfield  {journal} {\bibinfo
  {journal} {Astropart. Phys.}\ }\textbf {\bibinfo {volume} {60}},\ \bibinfo
  {pages} {72} (\bibinfo {year} {2015})}\BibitemShut {NoStop}%
\bibitem [{\citenamefont {Nam}(2016)}]{Nam:2015vak}%
  \BibitemOpen
  \bibfield  {author} {\bibinfo {author} {\bibfnamefont {J.}~\bibnamefont
  {Nam}},\ }\href {\doibase 10.22323/1.236.0663} {\bibfield  {journal}
  {\bibinfo  {journal} {PoS}\ }\textbf {\bibinfo {volume} {ICRC2015}},\
  \bibinfo {pages} {663} (\bibinfo {year} {2016})}\BibitemShut {NoStop}%
\bibitem [{\citenamefont {Vieregg}\ \emph {et~al.}(2016)\citenamefont
  {Vieregg}, \citenamefont {Bechtol},\ and\ \citenamefont
  {Romero-Wolf}}]{Vieregg:2015baa}%
  \BibitemOpen
  \bibfield  {author} {\bibinfo {author} {\bibfnamefont {A.}~\bibnamefont
  {Vieregg}}, \bibinfo {author} {\bibfnamefont {K.}~\bibnamefont {Bechtol}}, \
  and\ \bibinfo {author} {\bibfnamefont {A.}~\bibnamefont {Romero-Wolf}},\
  }\href {\doibase 10.1088/1475-7516/2016/02/005} {\bibfield  {journal}
  {\bibinfo  {journal} {JCAP}\ }\textbf {\bibinfo {volume} {02}},\ \bibinfo
  {pages} {005} (\bibinfo {year} {2016})},\ \Eprint
  {http://arxiv.org/abs/1504.08006} {arXiv:1504.08006 [astro-ph.IM]}
  \BibitemShut {NoStop}%
\bibitem [{\citenamefont {Sasaki}(2018)}]{Sasaki:2017uta}%
  \BibitemOpen
  \bibfield  {author} {\bibinfo {author} {\bibfnamefont {M.}~\bibnamefont
  {Sasaki}} (\bibinfo {collaboration} {NTA}),\ }\href {\doibase
  10.22323/1.301.0941} {\bibfield  {journal} {\bibinfo  {journal} {PoS}\
  }\textbf {\bibinfo {volume} {ICRC2017}},\ \bibinfo {pages} {941} (\bibinfo
  {year} {2018})}\BibitemShut {NoStop}%
\bibitem [{\citenamefont {Otte}(2019)}]{Otte:2018uxj}%
  \BibitemOpen
  \bibfield  {author} {\bibinfo {author} {\bibfnamefont {A.~N.}\ \bibnamefont
  {Otte}},\ }\href {\doibase 10.1103/PhysRevD.99.083012} {\bibfield  {journal}
  {\bibinfo  {journal} {Phys. Rev. D}\ }\textbf {\bibinfo {volume} {99}},\
  \bibinfo {pages} {083012} (\bibinfo {year} {2019})},\ \Eprint
  {http://arxiv.org/abs/1811.09287} {arXiv:1811.09287 [astro-ph.IM]}
  \BibitemShut {NoStop}%
\bibitem [{\citenamefont {Prohira}\ \emph {et~al.}(2020)\citenamefont {Prohira}
  \emph {et~al.}}]{Prohira:2019glh}%
  \BibitemOpen
  \bibfield  {author} {\bibinfo {author} {\bibfnamefont {S.}~\bibnamefont
  {Prohira}} \emph {et~al.},\ }\href {\doibase 10.1103/PhysRevLett.124.091101}
  {\bibfield  {journal} {\bibinfo  {journal} {Phys. Rev. Lett.}\ }\textbf
  {\bibinfo {volume} {124}},\ \bibinfo {pages} {091101} (\bibinfo {year}
  {2020})},\ \Eprint {http://arxiv.org/abs/1910.12830} {arXiv:1910.12830
  [astro-ph.HE]} \BibitemShut {NoStop}%
\bibitem [{\citenamefont {Wissel}\ \emph {et~al.}(2019)\citenamefont {Wissel},
  \citenamefont {Alvarez-Muñiz}, \citenamefont {Carvalho}, \citenamefont
  {Romero-Wolf}, \citenamefont {Schoorlemmer},\ and\ \citenamefont
  {Zas}}]{Wissel:2019alx}%
  \BibitemOpen
  \bibfield  {author} {\bibinfo {author} {\bibfnamefont {S.}~\bibnamefont
  {Wissel}}, \bibinfo {author} {\bibfnamefont {J.}~\bibnamefont
  {Alvarez-Muñiz}}, \bibinfo {author} {\bibfnamefont {W.~R.}\ \bibnamefont
  {Carvalho}}, \bibinfo {author} {\bibfnamefont {A.}~\bibnamefont
  {Romero-Wolf}}, \bibinfo {author} {\bibfnamefont {H.}~\bibnamefont
  {Schoorlemmer}}, \ and\ \bibinfo {author} {\bibfnamefont {E.}~\bibnamefont
  {Zas}},\ }\href {\doibase 10.1051/epjconf/201921604007} {\bibfield  {journal}
  {\bibinfo  {journal} {EPJ Web Conf.}\ }\textbf {\bibinfo {volume} {216}},\
  \bibinfo {pages} {04007} (\bibinfo {year} {2019})}\BibitemShut {NoStop}%
\bibitem [{\citenamefont {Wissel}\ \emph {et~al.}(2020)\citenamefont {Wissel}
  \emph {et~al.}}]{Wissel:2020sec}%
  \BibitemOpen
  \bibfield  {author} {\bibinfo {author} {\bibfnamefont {S.}~\bibnamefont
  {Wissel}} \emph {et~al.},\ }\href@noop {} {\  (\bibinfo {year} {2020})},\
  \Eprint {http://arxiv.org/abs/2004.12718} {arXiv:2004.12718 [astro-ph.IM]}
  \BibitemShut {NoStop}%
\bibitem [{\citenamefont {Romero-Wolf}\ \emph {et~al.}(2020)\citenamefont
  {Romero-Wolf} \emph {et~al.}}]{Romero-Wolf:2020pzh}%
  \BibitemOpen
  \bibfield  {author} {\bibinfo {author} {\bibfnamefont {A.}~\bibnamefont
  {Romero-Wolf}} \emph {et~al.},\ }in\ \href@noop {} {\emph {\bibinfo
  {booktitle} {{Latin American Strategy Forum for Research Infrastructure}}}}\
  (\bibinfo {year} {2020})\ \Eprint {http://arxiv.org/abs/2002.06475}
  {arXiv:2002.06475 [astro-ph.IM]} \BibitemShut {NoStop}%
\bibitem [{\citenamefont {Halzen}\ and\ \citenamefont
  {Saltzberg}(1998)}]{Halzen:1998be}%
  \BibitemOpen
  \bibfield  {author} {\bibinfo {author} {\bibfnamefont {F.}~\bibnamefont
  {Halzen}}\ and\ \bibinfo {author} {\bibfnamefont {D.}~\bibnamefont
  {Saltzberg}},\ }\href {\doibase 10.1103/PhysRevLett.81.4305} {\bibfield
  {journal} {\bibinfo  {journal} {Phys. Rev. Lett.}\ }\textbf {\bibinfo
  {volume} {81}},\ \bibinfo {pages} {4305} (\bibinfo {year} {1998})},\ \Eprint
  {http://arxiv.org/abs/hep-ph/9804354} {arXiv:hep-ph/9804354} \BibitemShut
  {NoStop}%
\bibitem [{\citenamefont {Dutta}\ \emph {et~al.}(2002)\citenamefont {Dutta},
  \citenamefont {Reno},\ and\ \citenamefont {Sarcevic}}]{Dutta:2002zc}%
  \BibitemOpen
  \bibfield  {author} {\bibinfo {author} {\bibfnamefont {S.~I.}\ \bibnamefont
  {Dutta}}, \bibinfo {author} {\bibfnamefont {M.~H.}\ \bibnamefont {Reno}}, \
  and\ \bibinfo {author} {\bibfnamefont {I.}~\bibnamefont {Sarcevic}},\ }\href
  {\doibase 10.1103/PhysRevD.66.077302} {\bibfield  {journal} {\bibinfo
  {journal} {Phys. Rev. D}\ }\textbf {\bibinfo {volume} {66}},\ \bibinfo
  {pages} {077302} (\bibinfo {year} {2002})},\ \Eprint
  {http://arxiv.org/abs/hep-ph/0207344} {arXiv:hep-ph/0207344} \BibitemShut
  {NoStop}%
\bibitem [{\citenamefont {Bigas}\ \emph {et~al.}(2008)\citenamefont {Bigas},
  \citenamefont {Deligny}, \citenamefont {Payet},\ and\ \citenamefont
  {Van~Elewyck}}]{Bigas:2008sw}%
  \BibitemOpen
  \bibfield  {author} {\bibinfo {author} {\bibfnamefont {O.~B.}\ \bibnamefont
  {Bigas}}, \bibinfo {author} {\bibfnamefont {O.}~\bibnamefont {Deligny}},
  \bibinfo {author} {\bibfnamefont {K.}~\bibnamefont {Payet}}, \ and\ \bibinfo
  {author} {\bibfnamefont {V.}~\bibnamefont {Van~Elewyck}},\ }\href {\doibase
  10.1103/PhysRevD.78.063002} {\bibfield  {journal} {\bibinfo  {journal} {Phys.
  Rev. D}\ }\textbf {\bibinfo {volume} {78}},\ \bibinfo {pages} {063002}
  (\bibinfo {year} {2008})},\ \Eprint {http://arxiv.org/abs/0806.2126}
  {arXiv:0806.2126 [astro-ph]} \BibitemShut {NoStop}%
\bibitem [{\citenamefont {Kusenko}\ and\ \citenamefont
  {Weiler}(2002)}]{Kusenko:2001gj}%
  \BibitemOpen
  \bibfield  {author} {\bibinfo {author} {\bibfnamefont {A.}~\bibnamefont
  {Kusenko}}\ and\ \bibinfo {author} {\bibfnamefont {T.~J.}\ \bibnamefont
  {Weiler}},\ }\href {\doibase 10.1103/PhysRevLett.88.161101} {\bibfield
  {journal} {\bibinfo  {journal} {Phys. Rev. Lett.}\ }\textbf {\bibinfo
  {volume} {88}},\ \bibinfo {pages} {161101} (\bibinfo {year} {2002})},\
  \Eprint {http://arxiv.org/abs/hep-ph/0106071} {arXiv:hep-ph/0106071}
  \BibitemShut {NoStop}%
\bibitem [{\citenamefont {Palomares-Ruiz}\ \emph {et~al.}(2006)\citenamefont
  {Palomares-Ruiz}, \citenamefont {Irimia},\ and\ \citenamefont
  {Weiler}}]{PalomaresRuiz:2005xw}%
  \BibitemOpen
  \bibfield  {author} {\bibinfo {author} {\bibfnamefont {S.}~\bibnamefont
  {Palomares-Ruiz}}, \bibinfo {author} {\bibfnamefont {A.}~\bibnamefont
  {Irimia}}, \ and\ \bibinfo {author} {\bibfnamefont {T.~J.}\ \bibnamefont
  {Weiler}},\ }\href {\doibase 10.1103/PhysRevD.73.083003} {\bibfield
  {journal} {\bibinfo  {journal} {Phys. Rev. D}\ }\textbf {\bibinfo {volume}
  {73}},\ \bibinfo {pages} {083003} (\bibinfo {year} {2006})},\ \Eprint
  {http://arxiv.org/abs/astro-ph/0512231} {arXiv:astro-ph/0512231} \BibitemShut
  {NoStop}%
\bibitem [{\citenamefont {Alvarez-Muñiz}\ \emph {et~al.}(2018)\citenamefont
  {Alvarez-Muñiz}, \citenamefont {Carvalho}, \citenamefont {Cummings},
  \citenamefont {Payet}, \citenamefont {Romero-Wolf}, \citenamefont
  {Schoorlemmer},\ and\ \citenamefont {Zas}}]{Alvarez-Muniz:2017mpk}%
  \BibitemOpen
  \bibfield  {author} {\bibinfo {author} {\bibfnamefont {J.}~\bibnamefont
  {Alvarez-Muñiz}}, \bibinfo {author} {\bibfnamefont {W.~R.}\ \bibnamefont
  {Carvalho}}, \bibinfo {author} {\bibfnamefont {A.~L.}\ \bibnamefont
  {Cummings}}, \bibinfo {author} {\bibfnamefont {K.}~\bibnamefont {Payet}},
  \bibinfo {author} {\bibfnamefont {A.}~\bibnamefont {Romero-Wolf}}, \bibinfo
  {author} {\bibfnamefont {H.}~\bibnamefont {Schoorlemmer}}, \ and\ \bibinfo
  {author} {\bibfnamefont {E.}~\bibnamefont {Zas}},\ }\href {\doibase
  10.1103/PhysRevD.97.023021, 10.1103/PhysRevD.99.069902} {\bibfield  {journal}
  {\bibinfo  {journal} {Phys. Rev.}\ }\textbf {\bibinfo {volume} {D97}},\
  \bibinfo {pages} {023021} (\bibinfo {year} {2018})},\ \bibinfo {note}
  {[erratum: Phys. Rev.D99,no.6,069902(2019)]},\ \Eprint
  {http://arxiv.org/abs/1707.00334} {arXiv:1707.00334 [astro-ph.HE]}
  \BibitemShut {NoStop}%
%%CITATION = ARXIV:1707.00334;%%
\bibitem [{\citenamefont {Alvarez-Muñiz}\ \emph {et~al.}(2017)\citenamefont
  {Alvarez-Muñiz}, \citenamefont {Carvalho}, \citenamefont {Cummings},
  \citenamefont {Payet}, \citenamefont {Romero-Wolf}, \citenamefont
  {Schoorlemmer},\ and\ \citenamefont {Zas}}]{nutausim}%
  \BibitemOpen
  \bibfield  {author} {\bibinfo {author} {\bibfnamefont {J.}~\bibnamefont
  {Alvarez-Muñiz}}, \bibinfo {author} {\bibfnamefont {W.~R.}\ \bibnamefont
  {Carvalho}}, \bibinfo {author} {\bibfnamefont {A.~L.}\ \bibnamefont
  {Cummings}}, \bibinfo {author} {\bibfnamefont {K.}~\bibnamefont {Payet}},
  \bibinfo {author} {\bibfnamefont {A.}~\bibnamefont {Romero-Wolf}}, \bibinfo
  {author} {\bibfnamefont {H.}~\bibnamefont {Schoorlemmer}}, \ and\ \bibinfo
  {author} {\bibfnamefont {E.}~\bibnamefont {Zas}},\ }\href@noop {} {}\bibinfo
  {howpublished} {\texttt{NuTauSim},
  \url{https://github.com/harmscho/NuTauSim}} (\bibinfo {year}
  {2017})\BibitemShut {NoStop}%
\bibitem [{\citenamefont {Safa}\ \emph {et~al.}(2020)\citenamefont {Safa},
  \citenamefont {Pizzuto}, \citenamefont {Argüelles}, \citenamefont {Halzen},
  \citenamefont {Hussain}, \citenamefont {Kheirandish},\ and\ \citenamefont
  {Vandenbroucke}}]{Safa:2019ege}%
  \BibitemOpen
  \bibfield  {author} {\bibinfo {author} {\bibfnamefont {I.}~\bibnamefont
  {Safa}}, \bibinfo {author} {\bibfnamefont {A.}~\bibnamefont {Pizzuto}},
  \bibinfo {author} {\bibfnamefont {C.~A.}\ \bibnamefont {Argüelles}},
  \bibinfo {author} {\bibfnamefont {F.}~\bibnamefont {Halzen}}, \bibinfo
  {author} {\bibfnamefont {R.}~\bibnamefont {Hussain}}, \bibinfo {author}
  {\bibfnamefont {A.}~\bibnamefont {Kheirandish}}, \ and\ \bibinfo {author}
  {\bibfnamefont {J.}~\bibnamefont {Vandenbroucke}},\ }\href {\doibase
  10.1088/1475-7516/2020/01/012} {\bibfield  {journal} {\bibinfo  {journal}
  {JCAP}\ }\textbf {\bibinfo {volume} {01}},\ \bibinfo {pages} {012} (\bibinfo
  {year} {2020})},\ \Eprint {http://arxiv.org/abs/1909.10487} {arXiv:1909.10487
  [hep-ph]} \BibitemShut {NoStop}%
\bibitem [{\citenamefont {Abramowicz}\ and\ \citenamefont
  {Levy}(1997)}]{Abramowicz:1997ms}%
  \BibitemOpen
  \bibfield  {author} {\bibinfo {author} {\bibfnamefont {H.}~\bibnamefont
  {Abramowicz}}\ and\ \bibinfo {author} {\bibfnamefont {A.}~\bibnamefont
  {Levy}},\ }\href@noop {} {\  (\bibinfo {year} {1997})},\ \Eprint
  {http://arxiv.org/abs/hep-ph/9712415} {arXiv:hep-ph/9712415 [hep-ph]}
  \BibitemShut {NoStop}%
%%CITATION = HEP-PH/9712415;%%
\bibitem [{\citenamefont {Martineau-Huynh}\ \emph {et~al.}(2016)\citenamefont
  {Martineau-Huynh} \emph {et~al.}}]{Martineau-Huynh:2015hae}%
  \BibitemOpen
  \bibfield  {author} {\bibinfo {author} {\bibfnamefont {O.}~\bibnamefont
  {Martineau-Huynh}} \emph {et~al.} (\bibinfo {collaboration} {GRAND}),\
  }\bibfield  {booktitle} {\emph {\bibinfo {booktitle} {{Proceedings, 7th Very
  Large Volume Neutrino Telescope Workshop (VLVnT 2015): Rome, Italy, September
  14-16, 2015}}},\ }\href {\doibase 10.1051/epjconf/201611603005} {\bibfield
  {journal} {\bibinfo  {journal} {EPJ Web Conf.}\ }\textbf {\bibinfo {volume}
  {116}},\ \bibinfo {pages} {03005} (\bibinfo {year} {2016})},\ \Eprint
  {http://arxiv.org/abs/1508.01919} {arXiv:1508.01919 [astro-ph.HE]}
  \BibitemShut {NoStop}%
%%CITATION = ARXIV:1508.01919;%%
\bibitem [{\citenamefont {Anchordoqui}\ \emph {et~al.}(2020)\citenamefont
  {Anchordoqui} \emph {et~al.}}]{Anchordoqui:2019omw}%
  \BibitemOpen
  \bibfield  {author} {\bibinfo {author} {\bibfnamefont {L.~A.}\ \bibnamefont
  {Anchordoqui}} \emph {et~al.},\ }\href {\doibase 10.1103/PhysRevD.101.023012}
  {\bibfield  {journal} {\bibinfo  {journal} {Phys. Rev.}\ }\textbf {\bibinfo
  {volume} {D101}},\ \bibinfo {pages} {023012} (\bibinfo {year} {2020})},\
  \Eprint {http://arxiv.org/abs/1907.03694} {arXiv:1907.03694 [astro-ph.HE]}
  \BibitemShut {NoStop}%
%%CITATION = ARXIV:1907.03694;%%
\bibitem [{\citenamefont {Dziewonski}\ and\ \citenamefont
  {Anderson}(1981)}]{Dziewonski:1981xy}%
  \BibitemOpen
  \bibfield  {author} {\bibinfo {author} {\bibfnamefont {A.}~\bibnamefont
  {Dziewonski}}\ and\ \bibinfo {author} {\bibfnamefont {D.}~\bibnamefont
  {Anderson}},\ }\href {\doibase 10.1016/0031-9201(81)90046-7} {\bibfield
  {journal} {\bibinfo  {journal} {Phys. Earth Planet. Interiors}\ }\textbf
  {\bibinfo {volume} {25}},\ \bibinfo {pages} {297} (\bibinfo {year}
  {1981})}\BibitemShut {NoStop}%
\bibitem [{\citenamefont {Bakhti}\ and\ \citenamefont
  {Smirnov}(2020)}]{Bakhti:2020tcj}%
  \BibitemOpen
  \bibfield  {author} {\bibinfo {author} {\bibfnamefont {P.}~\bibnamefont
  {Bakhti}}\ and\ \bibinfo {author} {\bibfnamefont {A.~Y.}\ \bibnamefont
  {Smirnov}},\ }\href {\doibase 10.1103/PhysRevD.101.123031} {\bibfield
  {journal} {\bibinfo  {journal} {Phys. Rev. D}\ }\textbf {\bibinfo {volume}
  {101}},\ \bibinfo {pages} {123031} (\bibinfo {year} {2020})},\ \Eprint
  {http://arxiv.org/abs/2001.08030} {arXiv:2001.08030 [hep-ph]} \BibitemShut
  {NoStop}%
\bibitem [{\citenamefont {Ahlers}\ \emph {et~al.}(2018)\citenamefont {Ahlers},
  \citenamefont {Denton},\ and\ \citenamefont {Rameez}}]{Ahlers:2017wpb}%
  \BibitemOpen
  \bibfield  {author} {\bibinfo {author} {\bibfnamefont {M.}~\bibnamefont
  {Ahlers}}, \bibinfo {author} {\bibfnamefont {P.}~\bibnamefont {Denton}}, \
  and\ \bibinfo {author} {\bibfnamefont {M.}~\bibnamefont {Rameez}},\ }\href
  {\doibase 10.22323/1.301.0282} {\bibfield  {journal} {\bibinfo  {journal}
  {PoS}\ }\textbf {\bibinfo {volume} {ICRC2017}},\ \bibinfo {pages} {282}
  (\bibinfo {year} {2018})}\BibitemShut {NoStop}%
\bibitem [{\citenamefont {Aab}\ \emph {et~al.}(2018)\citenamefont {Aab} \emph
  {et~al.}}]{Aab:2018chp}%
  \BibitemOpen
  \bibfield  {author} {\bibinfo {author} {\bibfnamefont {A.}~\bibnamefont
  {Aab}} \emph {et~al.} (\bibinfo {collaboration} {Pierre Auger}),\ }\href
  {\doibase 10.3847/2041-8213/aaa66d} {\bibfield  {journal} {\bibinfo
  {journal} {Astrophys. J. Lett.}\ }\textbf {\bibinfo {volume} {853}},\
  \bibinfo {pages} {L29} (\bibinfo {year} {2018})},\ \Eprint
  {http://arxiv.org/abs/1801.06160} {arXiv:1801.06160 [astro-ph.HE]}
  \BibitemShut {NoStop}%
\bibitem [{\citenamefont {Abbasi}\ \emph {et~al.}(2018)\citenamefont {Abbasi}
  \emph {et~al.}}]{Abbasi:2018tqo}%
  \BibitemOpen
  \bibfield  {author} {\bibinfo {author} {\bibfnamefont {R.}~\bibnamefont
  {Abbasi}} \emph {et~al.} (\bibinfo {collaboration} {Telescope Array}),\
  }\href {\doibase 10.3847/2041-8213/aaebf9} {\bibfield  {journal} {\bibinfo
  {journal} {Astrophys. J. Lett.}\ }\textbf {\bibinfo {volume} {867}},\
  \bibinfo {pages} {L27} (\bibinfo {year} {2018})},\ \Eprint
  {http://arxiv.org/abs/1809.01573} {arXiv:1809.01573 [astro-ph.HE]}
  \BibitemShut {NoStop}%
\bibitem [{\citenamefont {Aab}\ \emph {et~al.}(2020{\natexlab{a}})\citenamefont
  {Aab} \emph {et~al.}}]{Aab:2020xgf}%
  \BibitemOpen
  \bibfield  {author} {\bibinfo {author} {\bibfnamefont {A.}~\bibnamefont
  {Aab}} \emph {et~al.} (\bibinfo {collaboration} {Pierre Auger}),\ }\href
  {\doibase 10.3847/1538-4357/ab7236} {\bibfield  {journal} {\bibinfo
  {journal} {Astrophys. J.}\ }\textbf {\bibinfo {volume} {891}},\ \bibinfo
  {pages} {142} (\bibinfo {year} {2020}{\natexlab{a}})},\ \Eprint
  {http://arxiv.org/abs/2002.06172} {arXiv:2002.06172 [astro-ph.HE]}
  \BibitemShut {NoStop}%
\bibitem [{\citenamefont {Abbasi}\ \emph {et~al.}(2020)\citenamefont {Abbasi}
  \emph {et~al.}}]{Abbasi:2020fxl}%
  \BibitemOpen
  \bibfield  {author} {\bibinfo {author} {\bibfnamefont {R.}~\bibnamefont
  {Abbasi}} \emph {et~al.} (\bibinfo {collaboration} {Telescope Array}),\
  }\href@noop {} {\  (\bibinfo {year} {2020})},\ \Eprint
  {http://arxiv.org/abs/2005.07312} {arXiv:2005.07312 [astro-ph.HE]}
  \BibitemShut {NoStop}%
\bibitem [{\citenamefont {Aab}\ \emph {et~al.}(2020{\natexlab{b}})\citenamefont
  {Aab} \emph {et~al.}}]{Aab:2020mfn}%
  \BibitemOpen
  \bibfield  {author} {\bibinfo {author} {\bibfnamefont {A.}~\bibnamefont
  {Aab}} \emph {et~al.} (\bibinfo {collaboration} {Pierre Auger}),\ }\href@noop
  {} {\  (\bibinfo {year} {2020}{\natexlab{b}})},\ \Eprint
  {http://arxiv.org/abs/2004.10591} {arXiv:2004.10591 [astro-ph.HE]}
  \BibitemShut {NoStop}%
\bibitem [{\citenamefont {Ahlers}\ \emph {et~al.}(2016)\citenamefont {Ahlers},
  \citenamefont {Bai}, \citenamefont {Barger},\ and\ \citenamefont
  {Lu}}]{Ahlers:2015moa}%
  \BibitemOpen
  \bibfield  {author} {\bibinfo {author} {\bibfnamefont {M.}~\bibnamefont
  {Ahlers}}, \bibinfo {author} {\bibfnamefont {Y.}~\bibnamefont {Bai}},
  \bibinfo {author} {\bibfnamefont {V.}~\bibnamefont {Barger}}, \ and\ \bibinfo
  {author} {\bibfnamefont {R.}~\bibnamefont {Lu}},\ }\href {\doibase
  10.1103/PhysRevD.93.013009} {\bibfield  {journal} {\bibinfo  {journal} {Phys.
  Rev. D}\ }\textbf {\bibinfo {volume} {93}},\ \bibinfo {pages} {013009}
  (\bibinfo {year} {2016})},\ \Eprint {http://arxiv.org/abs/1505.03156}
  {arXiv:1505.03156 [hep-ph]} \BibitemShut {NoStop}%
\bibitem [{\citenamefont {Denton}\ \emph {et~al.}(2017)\citenamefont {Denton},
  \citenamefont {Marfatia},\ and\ \citenamefont {Weiler}}]{Denton:2017csz}%
  \BibitemOpen
  \bibfield  {author} {\bibinfo {author} {\bibfnamefont {P.~B.}\ \bibnamefont
  {Denton}}, \bibinfo {author} {\bibfnamefont {D.}~\bibnamefont {Marfatia}}, \
  and\ \bibinfo {author} {\bibfnamefont {T.~J.}\ \bibnamefont {Weiler}},\
  }\href {\doibase 10.1088/1475-7516/2017/08/033} {\bibfield  {journal}
  {\bibinfo  {journal} {JCAP}\ }\textbf {\bibinfo {volume} {08}},\ \bibinfo
  {pages} {033} (\bibinfo {year} {2017})},\ \Eprint
  {http://arxiv.org/abs/1703.09721} {arXiv:1703.09721 [astro-ph.HE]}
  \BibitemShut {NoStop}%
\bibitem [{\citenamefont {Aartsen}\ \emph
  {et~al.}(2017{\natexlab{b}})\citenamefont {Aartsen} \emph
  {et~al.}}]{Aartsen:2017ujz}%
  \BibitemOpen
  \bibfield  {author} {\bibinfo {author} {\bibfnamefont {M.}~\bibnamefont
  {Aartsen}} \emph {et~al.} (\bibinfo {collaboration} {IceCube}),\ }\href
  {\doibase 10.3847/1538-4357/aa8dfb} {\bibfield  {journal} {\bibinfo
  {journal} {Astrophys. J.}\ }\textbf {\bibinfo {volume} {849}},\ \bibinfo
  {pages} {67} (\bibinfo {year} {2017}{\natexlab{b}})},\ \Eprint
  {http://arxiv.org/abs/1707.03416} {arXiv:1707.03416 [astro-ph.HE]}
  \BibitemShut {NoStop}%
\end{thebibliography}%

\end{document}